\documentclass[aps,preprint]{revtex4}
\usepackage{amsfonts}
\usepackage{amsmath}
\usepackage{amssymb}
\usepackage{graphicx}

\providecommand{\U}[1]{\protect\rule{.1in}{.1in}}
\newcommand{\ba}{\begin{eqnarray}}
\newcommand{\ea}{\end{eqnarray}}
\def\beq{\begin{equation}}
\def\eeq{\end{equation}}
\newcommand{\form}[1]{(\ref{#1})}
\newcommand{\nm}{\nu_{\mu}}

\begin{document}

\begin{titlepage}
\begin{flushright}
CERN-TH-PH/2007-267\\
December 2007
\end{flushright}

\vspace*{1cm}

\begin{centering}
{\large {\bf Quantum-Gravity Decoherence Effects in Neutrino Oscillations:
Expected Constraints from CNGS and J-PARC}}

\vspace{1cm}

{\bf Nick~E.~Mavromatos}$^{a}$,~~{\bf Anselmo Meregaglia}$^{b,c}$,
~~{\bf Andr\' e Rubbia}$^b$, {\bf Alexander~S.~Sakharov}$^{b,d}$ and {\bf Sarben
Sarkar$^{a}$}

\end{centering}

\vspace{1cm}

\begin{centering}

{\bf Abstract}

\end{centering}

Quantum decoherence, the evolution of pure states into mixed states, may be a
feature of quantum-gravity models. In most cases, such models lead to fewer
neutrinos of all active flavours being
detected in a long baseline experiment as compared to
three-flavour standard neutrino oscillations.
We discuss the potential of the CNGS and J-PARC beams in constraining models of
quantum-gravity induced decoherence using neutrino oscillations as a probe.
We use as much as possible model-independent parameterizations, even though they are motivated by
specific microscopic models, for fits to the expected experimental data which yield bounds on
quantum-gravity decoherence parameters.

\vspace{2cm}
\begin{flushleft}
$^{a}$ Department of Physics, King's College London,
University of London, Strand,
London, WC2R 2LS, United Kingdom \\
$^{b}$ Swiss Institute of Technology ETH-Z\"urich, 8093, Z\"urich, Switzerland\\
$^{c}$ IPHC, Universit� Louis Pasteur, CNRS/IN2P3, Strasbourg, France\\
$^{d}$ CERN, Theory Division, Physics Department Geneva 23 CH 1211,
Switzerland\\
\end{flushleft}
\end{titlepage}

\section{Introduction\label{sec:1}}

If microscopic black holes, or other defects forming space-time foam, exist
in the vacuum state of quantum gravity (QG)~\cite{wheeler,hawking}, this
state, in our view, will constitute an \textquotedblleft
environment\textquotedblright\ which will be characterised by some \textit{%
entanglement entropy}, due to its interaction with low-energy matter. Based
on earlier work by Maldacena~\cite{maldacena} on anti-de-Sitter space black
holes Hawking~\cite{hawkingrecent} has recently claimed the absence of
information loss in quantum gravity. He argued that in a superposition of
topologies, the non-unitary contributions associated with the non-trivial
topology decay exponentially with time, leaving only contributions to the
path integral from the unitary topologically-trivial configurations. However
we do not believe that this settles the issue of induced decoherence in
quantum gravity since firstly the concept of anti-de Sitter space times as a
\textquotedblleft regulator\textquotedblright\ has been used and secondly a
Euclidean formulation of the path integral for quantum gravity has been
adopted. Both these features may not be shared by an eventual true theory of
quantum gravity. Furthermore, entanglement entropy is still present for an
outside observer~\cite{entanglementrecent} \ and space-time foam need not
consist only of black holes; other defects, e.g. point-like D-branes, might
be present. A stochastic fluctuation of populations of point-like
D-particles~\cite{horizons}, for instance, which are known to obey infinite
statistics~\cite{benattiinfstat} due to their infinite internal stringy
states, could constitute such a decoherening environment for matter
propagation. There are also other quantum-gravity issues that are not
completely understood. In particular, it is possible that the entire concept
of local effective Lagrangians breaks down in such situations.

The matter system in such a case behaves as an open quantum mechanical
system, exhibiting \emph{decoherence}, which has in principle detectable
experimental signatures. In the context of a phenomenological
parametrization of quantum-gravity induced decoherence the first tests along
these lines have been proposed in \cite{ehns} . A more microscopic
consideration was given in \cite{banks}, where the proposed parametrization
of decoherent effects of quantum gravity was forced to obey the Lindblad~%
\cite{lindblad,gorini} formalism of open systems, employing completely
positive dynamical semigroup maps. This latter phenomenology, however, may
not be a true feature of a quantum theory of gravity.

The decoherent approach to quantum gravity, entailing entanglement entropy,
has been followed by some of the authors~\cite{sarkar,mavromatos} in many
phenomenological tests or microscopic models of space-time foam~\cite%
{horizons}, within the framework of non-critical string theory; the latter
may be a viable (non-equilibrium) theory of space-time foam~\cite{emn},
based on an identification of time with the Liouville mode. The latter is
viewed as a dynamical local renormalization-group scale on the world-sheet
of a non-conformal string. The non-conformality of the string is the result
of its interaction with backgrounds which are out of equilibrium, such as
those provided by twinkling microscopic black holes in the foam. The entropy
in this case can be identified with the world-sheet conformal anomaly of a $%
\sigma$-model describing the propagation of a matter string in this
fluctuating background~\cite{emn}. Although within critical string theory,
arguments have been given that entanglement entropy can characterise the
number of microstates of Anti-de-Sitter black holes~\cite{entanglement}, we
do not find these to be entirely conclusive, and moreover an extension of
such counting to microscopic dynamical black holes, that characterise a
space time foam, is far from being understood. For instance, the process of
formation and annihilation of microscopic black holes or other singular
fluctuations in space time, including defects, may not be described by
critical string theory methods.

In view of the above issues, it is evident that the debate concerning
space-time foam remains open. The thermal aspects of an evaporating black
hole are suggestive that the environment due to quantum-gravity is a sort of
``thermal'' heat bath. This has been pursued by some authors, notably in
ref.~\cite{garay}. Another proposal, the D-particle foam model~\cite%
{horizons}, considers the gravitational fluctuations that could yield a
foamy structure of space-time to be D-particles (point-like stringy defects)
interacting with closed strings. There are no thermal aspects but there is
still the formation of horizons and entanglement entropy within a
fluctuating metric framework.

In general, for phenomenological purposes, the important feature of such
situations is the fact that gravitational environments, arising from
space-time foam or some other, possibly semi-classical feature of QG, can
still be described by non-unitary evolutions of a density matrix $\rho $.
Such equations have the form
\begin{equation}
\partial _{t}\rho =\Lambda _{1}\rho +\Lambda _{2}\rho
\end{equation}%
where
\begin{equation*}
\Lambda _{1}\rho =\frac{i}{\hbar }\left[ {\rho ,H}\right]  \label{ME}
\end{equation*}%
and $H$ is the hamiltonian with a stochastic element in a classical metric.
Such effects may arise from back-reaction of matter within a quantum theory
of gravity \cite{ehns,hu} which decoheres the gravitational state to give a
stochastic ensemble description. Furthermore within models of D-particle
foam arguments in favour of a stochastic metric have been given~\cite{sarkar}%
. The Liouvillian term $\Lambda _{2}\rho $ gives rise to a non-unitary
evolution. A common approach to $\Lambda _{2}\rho ,$ is to parametrise the
Liouvillian in a so called Lindblad form~\cite{lindblad,gorini} but this is
not based on microscopic physics. We note at this point that any non-linear
evolutions that may characterise a full theory of QG (see e.g. a
manifestation in Liouville strings~\cite{emnnl}), can be ignored to a first
approximation appropriate for the accuracy of contemporary experimental
probes of QG. Generically space-time foam and the back-reaction of matter on
the gravitational metric may be modelled as a randomly fluctuating
environment; formalisms for open quantum mechanical systems propagating in
such random media can thus be applied and lead to concrete experimental
predictions. The approach to these questions have to be phenomenological to
some degree since QG is not sufficiently developed at a non-perturbative
level.

One of the most sensitive probes of such stochastic quantum-gravity
phenomena are neutrinos~\cite{lisi,mavromatos,barenboim,barenboim2,
Benatti:2001fa,Benatti:2000ph,Brustein:2001ik,bmsw,lisib,lisi1}, and in
particular high-energy ones~\cite{winstanley,y}. It is the point of this
article to study decoherence induced by non-linear space-time foam
fluctuations as a subdominant effect in neutrino oscillations at CNGS and
J-PARC beams after giving an overview of the framework of decoherence
phenomena in neutrino experiments.

A linear decoherence simplified model, of Lindblad type~\cite{lindblad} has
been used for the fit, following earlier work in \cite{gago}. The model of
\cite{gago} involved a diagonal decoherence matrix, and in the fit of~\cite%
{barenboim2} (which we will refer to as I), decoherence was assumed to be
dominant \textit{only} in the antineutrino sector, in order to fit the LSND
results~\cite{lsnd} pointing towards significant ${\overline{\nu }}%
_{e}\rightarrow {\overline{\nu }}_{\mu }$ oscillations, but suppressed
oscillations in the particle sector. In this way in I a fit was made to a
three generation model with the LSND \textquotedblleft
anomalous\textquotedblright\ result, without introducing a sterile neutrino.
The strong CPT violation in the decoherence sector, allowed for an equality
of neutrino mass differences between the two sectors, so as to be in
agreement with atmospheric and solar neutrino data. However, the reader
should bear in mind that the LSND result has not been finally confirmed by
MiniBoone~\cite{mb}. Nevertheless, this does not detract from the motivation
to use a decoherence-based formalism to discuss quantum gravity effects for
neutrinos.

The particular choice of I, which yielded the best fit to all available
neutrino data involved mixed energy dependence for the (antineutrino-sector)
decoherence coefficients, some of which were proportional to the neutrino
energies $E$, while the rest had been inversely proportional to it, $\propto
1/E$. In I, the coefficients proportional to $1/E$ were interpreted as
describing ordinary matter effects, whilst the ones proportional to $E$ were
assumed to correspond to \textit{genuine} quantum-gravity effects, whose
increase with the energy of the probe was consistent with the fact that the
higher the (anti)neutrino energy the larger the back reaction effect on the
quantum space time, and hence the larger the decoherence. The strong
difference assumed in I between the decoherence coefficients of the particle
and antiparticle sectors, although not incompatible with a breakdown of CPT
at a fundamental level~\cite{mavromatos}, appears at first sight somewhat
curious; in fact it is unlike any other case of decoherence for other
sensitive particle probes, like neutral mesons, examined in the past~\cite%
{ehns,lopez}. There, the oscillations between particle and antiparticle
sectors, necessitate a common decoherence environment between mesons and
antimesons. If one accepts the Universality of gravity, then, the best fit
of I seems incompatible with this property. Moreover, there are two more
problematic points of the fit in I, which were already discussed in that
reference. The first point concerns the complete positivity of the model. In
\cite{gago} the diagonal form of the decoherence matrix, used in I, was
taken \textit{ad hoc}, without explicit mention of the necessary conditions
to guarantee complete positivity, as required by the Lindblad approach.
Furthermore, the particular choice of the best fit of I, did \textit{not}
lead to positive definite probabilities for the \textit{entire} regime of
the parameter space of the model, although the probabilities were positive
definite for the portion of the parameter space appropriate for the various
neutrino experiments used for the fit. Specifically, it was found in I, that
with the particular choice of the decoherence parameters in the
(antineutrino sector), one obtains positive-definite transition
probabilities for energies $E>\mathcal{O}(1~\mathrm{MeV})$. The second, and
more important point, is that the best fit of I is good for all the neutrino
experiments available at the time, but \textit{unfortunately} it could not
reproduce the spectral distortion observed by the KamLand experiment~\cite%
{kamland}, whose first results came out simultaneously with the results of
I. In \cite{bmsw} (which we will refer to as II) the above weakness was
rectified by requiring general conditions among the coefficients that
guarantee complete positivity in the \textit{entire} parameter space for the
three-generation simplified Lindblad linear model of decoherence of \cite%
{gago}, used in I. In fact it was shown in II that it was possible to choose
the coefficients in a way to give a consistent and excellent fit to all
available neutrino data including the spectral distortion seen by KamLand,
and the LSND results~\cite{lsnd} for the transition probabilities in the
antineutrino sector.

It has been argued that quantum decoherence could be an alternative
description of neutrino flavour transitions. Fits to data by different
experiments such as Super-Kamiokande~\cite{sk} and KamLand~\cite{kamland}
have been performed and these clearly disfavour a decoherence explanation
for neutrino oscillations. However, quantum decoherence may still be a
\textit{marginal} effect in addition to neutrino oscillations and could give
rise to damping factors in the transition probabilities reducing the number
of active neutrinos being detected in a long baseline experiment compared to
what is expected from the standard neutrino oscillation scenario.

\textbf{\ }Our article is organized as follows. In section \ref{sec:2}, we
describe the physics of the "quantum-gravitational analogues" of the MSW
effect and of foam models endowed with stochastic fluctuations of the
space-time metric background. In section \ref{sec:3}, we discuss decoherence
signatures in the neutrino oscillations and review the existing bounds on
these parameters using the available neutrino data, including those from
KamLand~\cite{kamland}, indicating spectral distortions. Then, in section %
\ref{sec:4}, we present the damping signatures and the associated fitting
functions, which might be due to either the "quantum-gravitational analogue"
of the MSW effect or the stochastic fluctuations of the space-time metric
background. We are careful to consider various stochastic models of foam,
which lead to different damping signatures, depending on the details of the
underlying characteristic distribution functions~\cite{alexandre}. In
section \ref{sec:cngs}, we estimate the sensitivity of CNGS and J-PARC
experiments to the parameters of quantum-gravitational decoherence entering
the set of the above-mentioned damping signatures. Finally, in section \ref%
{sec:discussion}, we compare the sensitivities estimated by means of bounds
obtained from data on atmospheric, solar and KamLand neutrino oscillations,
as well as the neutral kaon system, and discuss the possible relevance of
our results in guiding the construction of models of (the still elusive
theory of) Quantum Gravity.

\section{Theoretical Models for Quantum-Gravity Decoherence and Neutrinos
\label{sec:2}}

The picture we envisage is the following: there are several parallel
three-brane worlds, one of which represents our observable Universe,
embedded in a higher-dimensional bulk, in which only gravitational (closed)
string states propagate. On the brane world there are only open string
states propagating, representing ordinary matter, with their ends attached
on the hypersurface. Of course, there are also closed string states, either
propagating along the longitudinal brane directions, or crossing the brane
boundary from the bulk. As discussed in \cite{emw} consistent supersymmetric
models of D-particle foam can be constructed, in which the bulk space
between, say, two parallel brane worlds is populated by point-like
D-particle defects. Motion of either D-particles or branes, as required by
the need to have cosmological backgrounds for a brane observer, causes
supersymmetry breaking in both the brane and the bulk, and moreover results
in D-particles crossing the brane boundaries. These D-particle defects can
even represent compactified black holes from a four-dimensional view point,
with the extra dimensions being wrapped up appropriately in Planckian size
compactifications. One may then encounter a situation in which D-particle
point-like space time defects from the higher-dimensional bulk space time
cross the three brane (where ordinary matter resides), a radically different
picture from virtual excitations in a vacuum.

In ref. \cite{horizons} we have discussed the details of dynamical formation
of horizons on the brane world (in the context of (Liouville) string
theory), as a result of the encounter of brane matter with the crossing
D-particle defect. Schematically, ordinary string matter on the brane
creates - through back reaction (recoil) effects due to scattering off
D-particles- sufficient distortion of space-time for dynamical horizons,
surrounding the defect, to appear. The appearance of horizons in this way
looks - from the point of view of a four dimensional observer - as a
dynamical \textquotedblleft flashing on and off of a black
hole\textquotedblright , coming from the \textquotedblleft
vacuum\textquotedblright . Using (weak) positive energy conditions, we have
proven in \cite{horizons} that such configurations with horizons are
unstable. The life time of such objects is of the order of the Planck time,
since this is the time uncertainty for the defect to cross the brane world
and interact with stringy matter excitations. Once horizons form there is
entropy production and through this irreversibility and decoherence.
Consequently such stringy black hole defects are therefore not equivalent to
ordinary virtual particles in flat space-time field theories or in attempts
to discuss effective local quantum gravity approaches from the point of view
of decoherence (as those mentioned in \cite{kiefer}).

The presence of dynamical horizons is a real effect of the ground state of
quantum gravity (at least in such Liouville-string approaches to QG), which
implies "real\textquotedblright\ environmental entanglement of matter
systems with (gravitational) degrees of freedom behind the horizons. This
leads to the problem of loss of information for particles propagating
outside the horizon, and as such can lead to microscopic time
irreversibility \`{a} la Wald~\cite{wald}, and consequent CPT violation and
QG-induced decoherence. There is then a consequent non-unitary evolution of
particles outside the horizon. Somewhat general arguments (even in flat
space-times but with a boundary) have been put forward in the literature~%
\cite{srednicki} to justify this point of view. The general message of the
non-unitary evolution has then been extracted and codified with
phenomenological Lindblad master equations~\cite{lindblad,gorini} over two
decades~\cite{ehns,lopez,mavromatos} to describe particles evolving in
space-time foam.

We cannot, of course, advocate at this stage that this (non-critical,
Liouville) string approach, or similar, is the only consistent approach to
quantum gravity. Hence we by no means exclude the validity of the local
effective approach to QG, involving only virtual gravitons; in such cases
there might not be any induced decoherence~\cite{kiefer}, for reasons stated
above. It is therefore a challenging experimental issue to seek such
decoherence effects induced by quantum gravity, which would definitely
discriminate between several models of quantum gravity.

Moreover, there is another interesting possibility regarding neutrinos. As
pointed out recently in~\cite{barenboim}, the tiny mass differences between
neutrino flavours may themselves (in part) be the result of a CPT violating
quantum-gravity background. The phenomenon, if true, would be the
generalisation of the celebrated Mikheyev-Smirnov-Wolfenstein (MSW) effect~%
\cite{wolf,mikheev}. The latter arises from effective mass differences
between the various neutrino flavours, as a result of different type of
interactions of the various flavours with matter within the context of the
Standard Model. The phenomenon has been generalised to randomly fluctuating
media~\cite{loreti}, which are of relevance to solar and nuclear reactor $%
\beta $-decays neutrinos. This stochastic MSW effect will be more relevant
for us, since we consider space-time foam, as a random medium which induces
flavour-sensitive mass differences. If we can extrapolate~\cite{ms}
semi-classical results on black-hole evaporation, in both general relativity~%
\cite{gao} and string theory~\cite{lifschytz} to the quantum gravity foamy
ground state (assuming it exists and characterizes the ground state of some
(stochastic) quantum-gravity models, it follows that microscopic black holes
which are near extremal (and therefore electrically charged) would evaporate
significantly less, compared with their neutral counterparts. Thus, we may
assume~\cite{ms,barenboim}, that near extremal black holes in the foam would
\textquotedblleft live\textquotedblright\ longer, and as a result they would
have more time to interact with ordinary matter, such as neutrinos. Such
charged black holes would therefore constitute the dominant source of charge
fluctuations in the foam that could be responsible for foam-induced neutrino
mass differences according to the idea proposed in \cite{barenboim}. Indeed,
the emitted electrons from such black holes, which as stated above are
emitted preferentially, compared to muons or other charged particles, as
they are the lightest, would then have more time to interact (via coherent
standard model interactions) with the electron-neutrino currents, as opposed
to muon neutrinos. This would create a \textit{flavour bias} of the foam
medium, which could then be viewed~\cite{ms,barenboim} as the
\textquotedblleft quantum-gravitational analogue\textquotedblright\ of the
MSW effect~\cite{wolf,mikheev} in ordinary media (where, again, one has only
electrons, since the muons would decay quickly). In this sense, the quantum
gravity medium would be responsible for generating effective neutrino mass
differences~\cite{barenboim}. Since the charged-black holes lead to a
stochastically fluctuating medium, we will consider the formalism of the MSW
effect for stochastically fluctuating media~\cite{loreti}, where the density
of electrons would be replaced by the density of charged black hole/anti
black hole pairs.

\subsection{Quantum-Gravitational MSW effect and induced decoherence}

For simplicity, we will give theoretical details for the case of two
generations of neutrinos $\nu_{\mu}$ and $\nu_{\tau}$ with mass eigenvalues $%
m_1$ and $m_2$. We take the effective Hamiltonian to be of the form
\begin{equation}
H_{\mathrm{eff}}=H+n_{\mathrm{bh}}^{\mathrm{c}}(r)H_{I},
\label{EffectiveHamiltonian}
\end{equation}
where $H_{I}$ is a 2 $\times$ 2 matrix whose entries depend on the
interaction of the foam and neutrinos and $H$ is the free Hamiltonian. For
the purposes of this paper we take this matrix to be diagonal in flavour
space. \ Although we leave the entries as general constants, $a_{\nu_{i}}$,
we expect them to be of the form $\varpropto G_{N}n_{\mathrm{bh}}^{\mathrm{c}%
}(r)$; so we write $H_{I}$ as
\begin{equation}
H_{I}=\left(
\begin{array}{cc}
a_{\nu_{\mu}} & 0 \\
0 & a_{\nu_{\tau}}%
\end{array}
\right) .  \label{FlavourInt}
\end{equation}
where the foam medium is assumed to be described by Gaussian random
variables \cite{barenboim}. We take the average number of foam particles, $%
\langle n_{\mathrm{bh}}^{c}(t)\rangle=n_{0}$ (a constant), and $\langle
n_{bh}^{c}(t)n_{bh}^{c}(t^{\prime})\rangle\sim\Omega^{2}n_{0}^{2}%
\delta(t-t^{\prime })$. Following \cite{loreti} we can deduce the modified
time evolution of the density matrix as
\begin{equation}
\frac{\partial}{\partial t}\langle\rho\rangle=-i[H+n_{0}H_{I},\langle
\rho\rangle]-\Omega^{2}n_{0}^{2}[H_{I},[H_{I},\langle\rho\rangle ]]
\label{TwoFlavourMaster}
\end{equation}
where $\langle...\rangle$ represents the average over the random variables
of the foam. The double commutator is the CPT violating term since although
it \ is CP symmetric it induces time-irreversibility. It is also important
to note that $\Lambda_{2}$ here is of the Markovian-Liouville-Lindblad form
for a self-adjoint operator. This is as an appropriate form for decoherence
for environments about which we have little a priori knowledge. In the CPT
violating term we can require the density fluctuation parameter to be
different for the anti-particle sector from that for the particle sector,
i.e. $\bar{\Omega}\neq\Omega$, while keeping $\left\langle
n_{bh}^{c}(t)\right\rangle \equiv n_{0}$ the same in both sectors.
Physically this means that neutrinos and antineutrinos with the same
momenta, and hence interacting with the same amount of foam particles on
average, will evolve differently; this is a result of CPT violation.

We can expand the Hamiltonian and the density operator in terms of the Pauli
spin matrices $s_{\mu}$ (with $\frac{s_{0}}{2}=\mathbf{1}_{2}$ the $2\times2$
identity matrix) as follows
\begin{equation}
H_{\mathrm{eff}}=\sum_{\mu=0}^{3}(h_{\mu}+n_{0}h_{\mu}^{\prime})\frac{s_{\mu}%
}{2},\qquad\rho=\sum_{\nu=0}^{3}\rho_{\nu}\frac{s_{\nu}}{2}.
\label{expansion1}
\end{equation}
(where $H_{\mathrm{eff}}=H+n_{0}H_{I}$). We find that
\begin{equation}
h_{\mu}=\frac{m_{1}^{2}+m_{2}^{2}}{4k}\delta_{\mu0}+\frac{m_{1}^{2}-m_{2}^{2}%
}{2k}\delta_{\mu3}  \label{expansion2}
\end{equation}
and
\begin{equation}
n_{0}h_{\mu}^{\prime}=\frac{a_{\nu_{\mu}}+a_{\nu_{\tau}}}{2}%
\delta_{\mu0}+\left( a_{\nu_{\mu}}-a_{\nu_{\tau}}\right)
\sin2\theta\;\delta_{\mu1}+\left( a_{\nu_{\mu}}-a_{\nu_{\tau}}\right)
\cos2\theta\;\delta_{\mu3},  \label{expansion3}
\end{equation}
where $k$ is the neutrino energy scale.

\bigskip Adopting the Lindblad decoherence approach~\cite{lindblad}
described in~\cite{ms} one can arrive to expressions for the pure states
representing $\nu _{\mu }$ as given by
\begin{equation}
\left\langle \rho \right\rangle ^{\left( \nu _{\mu }\right) }=\frac{1}{2}%
\mathbf{1}_{2}+\sin \left( 2\theta \right) \frac{s_{1}}{2}+\cos \left(
2\theta \right) \frac{s_{3}}{2}  \label{Electron}
\end{equation}%
and the corresponding state for $\nu _{\tau }$ as
\begin{equation}
\left\langle \rho \right\rangle ^{\left( \nu _{\tau }\right) }=\frac{1}{2}%
\mathbf{1}_{2}-\sin \left( 2\theta \right) \frac{s_{1}}{2}-\cos \left(
2\theta \right) \frac{s_{3}}{2}.  \label{Muon}
\end{equation}%
If $\left\langle \rho \right\rangle \left( 0\right) =\left\langle \rho
\right\rangle ^{\left( \nu _{\mu }\right) }$ then the probability $P_{\nu
_{\mu }\rightarrow \nu _{\tau }}\left( t\right) $ of the transition $\nu
_{\mu }\rightarrow \nu _{\tau }$ is given by
\begin{equation}
P_{\nu _{\mu }\rightarrow \nu _{\tau }}\left( t\right) =Tr\left(
\left\langle \rho \right\rangle \left( t\right) \left\langle \rho
\right\rangle ^{\left( \nu _{\tau }\right) }\right) .  \label{probability}
\end{equation}%
In order to study decoherence we will calculate the eigenvectors $%
\overrightarrow{\mathfrak{e}}^{\left( _{i}\right) }$ and corresponding
eigenvalues $\mathfrak{\lambda }_{i}$ of \ $\mathcal{L}$ to leading order in
$\Omega ^{2}$. In terms of auxiliary variables $\mathcal{U}$ and $\mathcal{W}
$ where
\begin{equation}
\mathcal{U}=\left( a_{\nu _{\mu }}-a_{\nu _{\tau }}\right) \cos \left(
2\theta \right) +\frac{m_{1}^{2}-m_{2}^{2}}{2k}  \label{aux1}
\end{equation}%
and
\begin{equation}
\mathcal{W}=\left( a_{\nu _{\mu }}-a_{\nu _{\tau }}\right) \sin \left(
2\theta \right) ,  \label{aux2}
\end{equation}%
it is straightforward to show that
\begin{align}
\overrightarrow{\mathfrak{e}}^{_{^{\left( 1\right) }}}& \simeq \left( \frac{%
\mathcal{W}}{\mathcal{U}},0,1\right) ,  \notag \\
\overrightarrow{\mathfrak{e}}^{_{\left( 2\right) }}& \simeq \left( -\frac{%
\mathcal{U}}{\mathcal{W}},-i\frac{\sqrt{\mathcal{U}^{2}+\mathcal{W}^{2}}}{%
\mathcal{W}},1\right) ,  \label{eigenvect} \\
\overrightarrow{\mathfrak{e}}^{_{\left( 3\right) }}& \simeq \left( -\frac{%
\mathcal{U}}{\mathcal{W}},i\frac{\sqrt{\mathcal{U}^{2}+\mathcal{W}^{2}}}{%
\mathcal{W}},1\right) ,  \notag
\end{align}%
and%
\begin{align}
\mathfrak{\lambda }_{1}& \simeq -\Omega ^{2}\left( \mathcal{W}\cos \left(
2\theta \right) -\mathcal{U}\sin \left( 2\theta \right) \right) ^{2},  \notag
\\
\mathfrak{\lambda }_{2}& \simeq -i\sqrt{\mathcal{U}^{2}+\mathcal{W}^{2}}-%
\frac{\Omega ^{2}}{2}\left( \mathcal{U}^{2}+\mathcal{W}^{2}+\left( \mathcal{U%
}\cos \left( 2\theta \right) +\mathcal{W}\sin \left( 2\theta \right) \right)
^{2}\right) ,  \label{eigenval} \\
\mathfrak{\lambda }_{3}& \simeq i\sqrt{\mathcal{U}^{2}+\mathcal{W}^{2}}-%
\frac{\Omega ^{2}}{2}\left( \mathcal{U}^{2}+\mathcal{W}^{2}+\left( \mathcal{U%
}\cos \left( 2\theta \right) +\mathcal{W}\sin \left( 2\theta \right) \right)
^{2}\right) .  \notag
\end{align}%
The vector $\overrightarrow{\rho }$ $\left( 0\right) $ can be decomposed~%
\cite{ms} as
\begin{equation}
\overrightarrow{\rho }\left( 0\right) =\mathsf{b}_{1}\overrightarrow{%
\mathfrak{e}}^{_{^{\left( 1\right) }}}+\mathsf{b}_{2}\overrightarrow{%
\mathfrak{e}}^{_{^{\left( 2\right) }}}+\mathsf{b}_{2}\overrightarrow{%
\mathfrak{e}}^{_{^{\left( 3\right) }}}  \label{decomp}
\end{equation}%
with
\begin{equation}
\mathsf{b}_{1}=\frac{\mathcal{U}^{2}\cos \left( 2\theta \right) +\mathcal{UW}%
\sin \left( 2\theta \right) }{\mathcal{U}^{2}+\mathcal{W}^{2}}
\end{equation}%
and
\begin{equation}
\mathsf{b}_{2}=\frac{\mathcal{W}^{2}\cos \left( 2\theta \right) -\mathcal{UW}%
\sin \left( 2\theta \right) }{2\left( \mathcal{U}^{2}+\mathcal{W}^{2}\right)
}.
\end{equation}%
Hence
\begin{equation}
\rho \left( t\right) =\frac{1}{2}\left( \mathsf{b}_{1}e^{\lambda _{1}t}%
\overrightarrow{\mathfrak{e}}^{_{^{\left( 1\right) }}}.\overrightarrow{s}+%
\mathsf{b}_{2}\overrightarrow{\mathfrak{e}}^{_{^{\left( 2\right) }}}.%
\overrightarrow{s}+\mathsf{b}_{2}\overrightarrow{\mathfrak{e}}^{_{^{\left(
3\right) }}}.\overrightarrow{s}+\mathbf{1}_{2}\right) .
\end{equation}%
From this one can obtain from a standard analysis~\cite%
{Benatti:2001fa,Benatti:2000ph,loreti,bmsw} the following expression for the
neutrino transition probability $\nu _{\mu }\leftrightarrow \nu _{\tau }$ in
this case, to leading order in the small parameter $\Omega ^{2}\ll 1$:
{\small
\begin{align}
& P_{\nu _{\mu }\rightarrow \nu _{\tau }}=  \notag \\
& \frac{1}{2}+e^{-\Delta a_{\mu \tau }^{2}\Omega ^{2}t(1+\frac{\Delta
_{12}^{2}}{4\Gamma }(\cos (4\theta )-1))}\sin (t\sqrt{\Gamma })\sin
^{2}(2\theta )\Delta a_{\mu \tau }^{2}\Omega ^{2}\Delta _{12}^{2}\left(
\frac{3\sin ^{2}(2\theta )\Delta _{12}^{2}}{4\Gamma ^{5/2}}-\frac{1}{\Gamma
^{3/2}}\right)  \notag \\
& -e^{-\Delta a_{\mu \tau }^{2}\Omega ^{2}t(1+\frac{\Delta _{12}^{2}}{%
4\Gamma }(\cos (4\theta )-1))}\cos (t\sqrt{\Gamma })\sin ^{2}(2\theta )\frac{%
\Delta _{12}^{2}}{2\Gamma }  \notag \\
& -e^{-\frac{\Delta a_{\mu \tau }^{2}\Omega ^{2}t\Delta _{12}^{2}\sin
^{2}(2\theta )}{\Gamma }}\frac{(\Delta a_{\mu \tau }+\cos (2\theta )\Delta
_{12})^{2}}{2\Gamma }  \label{2genprob}
\end{align}%
} where $\Gamma =(\Delta a_{\mu \tau }\cos (2\theta )+\Delta
_{12})^{2}+\Delta a_{\mu \tau }^{2}\sin ^{2}(2\theta )~,$ $\Delta _{12}=%
\frac{\Delta m_{12}^{2}}{2k}$~and $\Delta a_{\mu \tau }\equiv a_{\nu _{\mu
}}-a_{\nu _{\tau }}$.

{}From (\ref{2genprob}) we easily conclude that the exponents of the damping
factors due to the stochastic-medium-induced decoherence, are of the generic
form, for $t=L$, with $L$ the oscillation length (in units of $c=1$):
\begin{equation*}
\mathrm{exponent}\sim-\Delta a_{\mu\tau}^{2}\Omega^{2}tf(\theta)~;~f(\theta
)=1+\frac{\Delta_{12}^{2}}{4\Gamma}(\cos(4\theta)-1)~,~\mathrm{or}~\frac{%
\Delta_{12}^{2}\sin^{2}(2\theta)}{\Gamma}
\end{equation*}
that is proportional to the stochastic fluctuations of the density of the
medium. The reader should note at this stage that, in the limit $\Delta
_{12}\rightarrow0$, which could characterize the situation in \cite%
{barenboim}, where the space-time foam effects on the induced neutrino mass
difference are the dominant ones, the damping factor is of the form $\mathrm{%
exponent}_{\mathrm{gravitational~MSW}}\sim-\Omega^{2}(\Delta
a_{\mu\tau})^{2}L~,$ with the precise value of the mixing angle $\theta$ not
affecting the leading order of the various exponents. However, in that case,
as follows from (\ref{2genprob}), the overall oscillation probability is
suppressed by factors proportional to $\Delta_{12}^{2}$, and, hence, the
stochastic gravitational MSW effect~\cite{barenboim}, although in principle
capable of inducing mass differences for neutrinos, however does not suffice
to produce the bulk of the oscillation probability, which is thus attributed
to conventional flavour physics. The damping exponent should then be
independent of the mixing angle for consistency. Indeed, we find the purely
gravitational MSW to give $\mathrm{exponent}_{\mathrm{gravitationalMSW}%
}\propto\Omega^{2}\Delta^{2}L$ which is independent of $\theta$. However,
this stochastic gravitational MSW effect, although capable of inducing
neutrino mass differences, gives an oscillation probability which is
suppressed by factors proportional to $\Delta_{12}^{2}$. Hence, the bulk of
the oscillation is due to conventional flavour physics.

After this theoretical discussion we now proceed to give a brief description
of the most important phenomenological consequences of such a scenario
involving decoherence. These can help in imposing stringent constraints on
the percentage of the neutrino mass difference that could be due to the
quantum-gravity medium. For simplicity we restrict ourselves to two
generations, which suffices for a demonstration of the important generic
properties of decoherence. The extension to three generations is
straightforward, albeit mathematically more complex~\cite{bmsw}.

We note here that, for gravitationally-induced MSW effects (due to, say,
black-hole foam models as in \cite{barenboim,ms})
\begin{equation*}
\Delta a_{\mu \tau }\propto G_{N}n_{0}
\end{equation*}%
with $G_{N}=1/M_{P}^{2}$, $M_{P}\sim 10^{19}~\mathrm{GeV}$, the
four-dimensional Planck scale, and in the case of the gravitational MSW-like
effect~\cite{barenboim} $n_{0}$ represents the density of charge black
hole/anti-black hole pairs. This gravitational coupling replaces the weak
interaction Fermi coupling constant $G_{F}$ in the conventional MSW effect.
This is the case that is relevant for this work. In such a situation the
density fluctuations $\Omega ^{2}$ can be assumed small compared to other
quantities present in the above formulae, and an expansion to leading order
in $\Omega ^{2}$ is appropriate.

\subsection{Stochastic fluctuations of Space-Time metric backgrounds}

There are other models of stochastic space-time foam also inducing
decoherence, for instance the ones discussed in~\cite{ms,bmsw}, in which one
averages over random (Gaussian) fluctuations of the background space-time
metric over which the neutrino propagates. In such an approach, one
considers merely the Hamiltonian of the neutrino in a stochastic metric
background. The stochastic fluctuations of the metric would then pertain to
the Hamiltonian (commutator) part of the density-matrix evolution. In
parallel, of course, one should also consider environmental
decoherence-interactions of Lindblad (or other) type, which would co-exist
with the decoherence effects due to the stochastic metric fluctuations in
the Hamiltonian. For definiteness in what follows we restrict ourselves only
to the Hamiltonian part, with the aim of demonstrating clearly the pertinent
effect and study their difference from Lindblad decoherence.

In this case, one obtains transition probabilities with exponential damping
factors in front of the oscillatory terms, but now the scaling with the
oscillation length (time) is quadratic~\cite{ms,bmsw}, consistent with time
reversal invariance of the neutrino Hamiltonian. For instance, for the two
generation case, which suffices for our qualitative purposes in this work,
we may consider stochastically fluctuating space-times with metrics
fluctuating along the direction of motion (for simplicity)~\cite{ms}
\begin{equation*}
g^{\mu\nu}=\left(
\begin{array}{cc}
-(a_{1}+1)^{2}+a_{2}^{2} & -a_{3}(a_{1}+1)+a_{2}(a_{4}+1) \\
-a_{3}(a_{1}+1)+a_{2}(a_{4}+1) & -a_{3}^{2}+(a_{4}+1)^{2}%
\end{array}
\right) .
\end{equation*}
with random variables $\langle a_{i}\rangle=0$ and $\langle
a_{i}a_{j}\rangle=\delta_{ij}\sigma_{i}$.

Two generation Dirac neutrinos, then, which are considered for definiteness
in \cite{ms} (one would obtain similar results, as far as decoherence
effects are concerned in the Majorana case), with an MSW interaction $V$ (of
unspecified origin, which thus could be a space-time foam effect) yield the
following oscillation probability from an initial state of flavour $1$ to $2$
is
\begin{equation}
\mathrm{Prob}(1\rightarrow2)=\sum\limits_{j,l}{U_{1j}}U_{2j}^{\ast}U_{1l}^{%
\ast}U_{2l}e^{i\left( {\omega}_{l}{-\omega_{j}}\right) t}
\end{equation}
where the time dependent part is
\begin{equation*}
U_{12}U_{22}^{\ast}U_{11}^{\ast}U_{21}e^{i\left( {\omega_{1}-\omega_{2}}%
\right) t}+U_{11}U_{21}^{\ast}U_{12}^{\ast}U_{22}e^{i\left( {\omega
_{2}-\omega_{1}}\right) t}
\end{equation*}
with $U$ the mixing matrix, which, in the two-flavour-dominance scenario we
are working on here for the sake of brevity, can be parametrised by a mixing
angle $\theta$:
\begin{equation}
U=\left(
\begin{array}{cc}
\cos\theta & \sin\theta \\
-\sin\theta & \cos\theta%
\end{array}
\right)
\end{equation}
Since the $\{a_{i}\}$ are assumed to be independent Gaussian variables, the
pertinent covariance matrix $\Xi$ has the diagonal form
\begin{equation}
\Xi=\left(
\begin{array}{cccc}
\frac{1}{\sigma_{1}} & 0 & 0 & 0 \\
0 & \frac{1}{\sigma_{2}} & 0 & 0 \\
0 & 0 & \frac{1}{\sigma_{3}} & 0 \\
0 & 0 & 0 & \frac{1}{\sigma_{4}}%
\end{array}
\right) ,
\end{equation}
with $\sigma_{i}>0$. The calculation of transition probabilities requires
the evaluation of the following average over the stochastic space-time
fluctuations $a_{i}$
\begin{equation}
\langle e^{i(\omega_{1}-\omega_{2})t}\rangle\equiv\int d^{4}a\exp(-\vec {a}%
\cdot\Xi\cdot\vec{a})e^{i(\omega_{1}-\omega_{2})t}\frac{\det\Xi}{\pi^{2}}.
\end{equation}
The result is~\cite{ms}
\begin{align}
& \langle e^{i(\omega_{1}-\omega_{2})t}\rangle= e^{i\frac{{\left( {z_{0}^{+}
- z_{0}^{-} } \right) t}}{k}} e^{-\frac{1}{2}\left( -i\sigma_{1} t\left(
\frac{(m_{1}^{2}-m_{2}^{2})}{k}+ V\cos2\theta\right) \right) } \times  \notag
\\
& e^{-\frac{1}{2}\left( \frac{i\sigma_{2}t}{2}\left( \frac{%
(m_{1}^{2}-m_{2}^{2})}{k}+V\cos2\theta\right) -\frac{i\sigma_{3}t}{2}V\cos
2\theta\right) } \times  \notag \\
& e^{-\left( \frac{(m_{1}^{2}-m^{2}_{2})^{2}}{2k^{2}} (9\sigma_{1}+\sigma
_{2}+\sigma_{3}+\sigma_{4})+\frac{2V\cos2\theta(m_{1}^{2}-m_{2}^{2})}{k}
(12\sigma_{1}+2\sigma_{2}-2\sigma_{3})\right) t^{2}}  \label{gravstoch}
\end{align}
where again $k$ is the neutrino energy scale, $\sigma_{i}~, i=1, \dots4$
parametrise appropriately the stochastic fluctuations of the metric in the
model of \cite{ms}, $\Upsilon= \frac{{Vk}}{{m_{1}^{2} - m_{2}^{2} }}$, $%
\left| \Upsilon\right| \ll1$, and $k^{2} \gg m_{1}^{2}~,~m_{2}^{2} $, and
\begin{align}
z_{0}^{+} & = \frac{1}{2}\left( m_{1}^{2} + \Upsilon(1 + \cos2\theta
)(m_{1}^{2} - m_{2}^{2} ) + \Upsilon^{2} (m_{1}^{2} - m_{2}^{2} )\sin^{2}
2\theta\right)  \notag \\
z_{0}^{-} & = \frac{1}{2}\left( m_{2}^{2} + \Upsilon(1 - \cos2\theta
)(m_{1}^{2} - m_{2}^{2} ) - \Upsilon^{2} (m_{1}^{2} - m_{2}^{2} )\sin^{2}
2\theta\right) ~.
\end{align}
Note that the metric fluctuations-$\sigma_{i}$ induced modifications of the
oscillation period, as well as exponential $e^{-(...)t^{2}}$ time-reversal
invariant damping factors~\cite{ms}, in contrast to the Lindblad
decoherence, in which the damping was of the form $e^{-(...)t}$. At first
thought, one may attribute this feature to the fact that, in this approach,
only the Hamiltonian terms are taken into account (in a stochastically
fluctuating metric background), and as such time reversal invariance $t
\to-t $ is not broken explicitly. But there is of course decoherence, and
the associated damping.

However, upon closer inspection things are not as simple. As shown in \cite%
{alexandre}, the power of the time variable $t$ in the associated damping is
\emph{crucially dependent } on the type of the distribution characterising
the gravitational fluctuations. In terms of our D-particle-recoil induced
stochastic model of space-time foam~\cite{horizons,sarkar,ms}, such
distributions characterise the ensemble of velocities of the gas of
D-particles involved in the foam, with which the neutrinos interact, and
which in turn affects the induced metric fluctuations, as explained above.
Assume, for instance, that the distribution of the (induced) metric
fluctuations are of Cauchy-Lorentz type, which could be induced by a
distribution of D-particle velocities in a D-foam model of the type
considered in \cite{emw}. Such a distribution has undefined mean and
variance as well as undefined or infinite higher moments. Assuming for
concreteness a case with zero mean ( in the sense of principal values), the
pertinent distribution function is taken to be \cite{alexandre}:
\begin{equation}
f(x)=\frac{\xi }{x^{2}+\xi ^{2}}  \label{cauchy}
\end{equation}%
with $\xi >0$ the characteristic scale parameter of the Cauchy-Lorentz
distribution. In that case, the pertinent statistical average of the
associated oscillation probability for a two-flavour neutrino problem would
be given, to leading order in an expansion in powers of $m_{i}/k$, of
interest to us here, by:
\begin{equation}
\langle e^{i(\omega _{1}-\omega _{2})t}\rangle _{\mathrm{CL}}\simeq \mathrm{%
exp}\left( ikt\Delta -\xi kt|\Delta |\right) ~,\qquad \Delta =\frac{%
m_{1}^{2}-m_{2}^{2}}{2k^{2}}  \label{CLdistribution}
\end{equation}%
From the above equation we do observe a linear damping, similar to the
Lindblad environment case. The important point to notice is that in this
case, the damping exponent is of order
\begin{equation}
\mathrm{exp}\left( -\Omega _{\mathrm{CL}}^{2}t\right) ~,~~\quad \mathrm{with}%
~~\quad \Omega _{\mathrm{CL}}^{2}\simeq \xi \frac{|m_{1}^{2}-m_{2}^{2}|}{2k}
\label{dampingCL}
\end{equation}%
The reader should compare the linear power of the (small) quantity $%
|m_{1}^{2}-m_{2}^{2}|/k$ entering the damping exponent (\ref{dampingCL}) in
the Cauchy-Lorentz stochastic foam model to the quadratic power of that
quantity entering the Gaussian model of foam (\ref{gravstoch}), where the
pertinent exponents are proportional to factors of the form $%
(m_{1}^{2}-m_{2}^{2})^{2}/k^{2}$ and hence \emph{much smaller}, provided of
course the parameters $\sigma ^{2}$ and $\xi $ are of \emph{similar order}.
At this stage, these are treated as \emph{phenomenological } parameters,
since their order depends on the details of the underlying model. For
instance, in the case of the D-particle foam model~\cite{emw,horizons}, such
an information depends on the dynamics of the gas of bulk D-particles, which
probably is an issue that can only be resolved within a microscopic M-theory
model.

At this stage we would like to draw the reader's attention to a
possible
interpretation~\cite{y} of the Lindblad-type exponential
damping~\form{dampingCL}
of the Cauchy-Lorentz distribution with $1/E$ dependence as a neutrino decay,
\beq
\label{lifetime}
\exp (-\Omega_{\rm CL}^2t)\equiv\exp (-t/\tau_{\rm lab})=\exp
 (-tm_{\nu_i}/E\tau_{\rm rest}),
\eeq
from which we can get a
lower limit on the (unstable) neutrinos life times. We shall
come back to this
issue in the concluding section of the article, when we provide the relevant
experimental bounds.

It must be stressed at this point, before closing this subsection, that the
above considerations, especially the ones concerning the form and order of
the decoherence damping factors, of interest to us in this work, although
derived in a two-dimensional toy-model (considering metric deformations
primarily along the direction of motion of the neutrino probe), nevertheless
are valid qualitatively in a full fledged four space time dimensional model.
This has been demonstrated in \cite{alexandre}, where realistic models of
neutrinos propagating in four-dimensional, stochastically fluctuating,
space-time backgrounds have been considered in detail, with results similar
to the ones considered in \cite{sarkar} and reviewed above.

\subsection{Mimicking Decoherence via Conventional Uncertainties in Neutrino
Energy and Oscillation Length}

A few remarks are now in order regarding the similarity of this latter type
of decoherence (\ref{gravstoch}) with the one mimicked~\cite{ohlsson} by
ordinary uncertainties in neutrino experiments over the precise energy $E$
of the beam (and in some cases over the oscillation length $L$). Indeed,
consider the Gaussian average of a generic neutrino oscillation probability
over the L/E dependence $\langle P \rangle= \int_{-\infty}^{\infty} dx P(x)
\frac{1}{\sigma\sqrt{2\pi}}e^{-\frac{(x-l)^{2}}{2\sigma^{2}}}~,$ with $%
l=\langle x \rangle$ and $\sigma=\sqrt{\langle(x-\langle x
\rangle)^{2}\rangle}$, $x=\frac{L}{4E}$, and assuming the independence of $L$
and $E$, which allows to write $\langle L/E\rangle=\langle L\rangle/\langle
E\rangle$. A pessimistic and an optimistic upper bound for $\sigma$ are
given by~\cite{ohlsson}

\begin{itemize}
\item pessimistic: $\sigma\simeq\Delta x =\Delta\frac{L}{4E} \leq\Delta L %
\big{|}\frac{\partial x}{\partial L}\big{|}_{L=\langle L\rangle, E=\langle E
\rangle}+\Delta E \big{|}\frac{\partial x} {\partial E}\big{|}_{L=\langle
L\rangle, E=\langle E \rangle}$

$=\frac{\langle L\rangle}{4\langle E\rangle} \left( \frac{\Delta L}{\langle
L\rangle}+\frac{\Delta E} {\langle E\rangle}\right) $

\item optimistic: $\sigma\le\frac{\langle L\rangle}{4\langle E\rangle} \sqrt{%
\left( \frac{\Delta L}{\langle L\rangle}\right) ^{2}+\left( \frac{\Delta E} {%
\langle E\rangle}\right) ^{2}} $
\end{itemize}

Then, it is easy to arrive at the expression~\cite{ohlsson} {\small
\begin{align}
& \langle P_{\alpha\beta} \rangle= \delta_{\alpha\beta} -  \notag \\
& 2 \sum_{a=1}^{n}\sum_{b=1, a<b}^{n}\mathrm{Re}\left( U_{\alpha a}^{*}
U_{\beta a}U_{\alpha b}U_{\beta b}^{*}\right) \left( 1 - \mathrm{cos}%
(2\ell\Delta m_{ab}^{2}) e^{-2\sigma^{2}(\Delta m_{ab}^{2})^{2}}\right)
\notag \\
& -2 \sum_{a=1}^{n} \sum_{b=1, a<b}^{n} \mathrm{Im}\left( U_{\alpha a}^{*}
U_{\beta a}U_{\alpha b}U_{\beta b}^{*}\right) \mathrm{sin}(2\ell\Delta
m_{ab}^{2}) e^{-2\sigma^{2}(\Delta m_{ab}^{2})^{2}}  \notag \\
\mathrm{with} \qquad\ell\equiv\frac{\langle L \rangle}{4\langle E \rangle }
\label{uncert}
\end{align}
} with $U$ the appropriate mixing matrix. Notice the $\sigma^{2}$ damping
factor of neutrino oscillation probabilities, which has the similar form in
terms of the oscillation length dependence ($L^{2}$ dependence) as the
corresponding damping factors due to the stochasticity of the space-time
background in (\ref{gravstoch}). It is noted, however, that here $l$ has to
do with the sensitivity of the experiment, and thus the physics is entirely
different.

In the case of space-time stochastic backgrounds, one could still have
induced uncertainties in $E$ and $L$, which however are of fundamental
origin, and are expected to be more suppressed than the uncertainties due to
ordinary physics, described above. Apart from their magnitude, their main
difference from the uncertainties in (\ref{uncert}) has to do with the
specific dependence of the corresponding $\sigma^{2}$ in that case on both $%
E $ and $L$. For generic space-time foam models it is expected that an
uncertainty in $E$ or $L$ due to the ``fuzziness'' of space time at a
fundamental (Planckian) level will increase with the energy of the probe, $%
\delta E/E, \delta L/L \propto (E/M_{P})^{\alpha}$, $\alpha> 0$, since the
higher the energy the bigger the disturbance (and hence back reaction)on the
space time medium. In contrast, ordinary matter effects decrease with the
energy of the probe~\cite{ohlsson,bow}.

\section{Previous Decoherent Fits with Existing Neutrino Data and Physical
Interpretation\label{sec:3}}

The first complete phenomenological attempt to fit decoherence models to
atmospheric neutrino data was done in \cite{lisi}, where for simplicity a
two-generation neutrino model with completely positive Lindblad decoherence,
characterised by a single parameter $\gamma$, and leading to exponential
damping with time of the relevant oscillatory terms in the respective
oscillation probabilities, was considered.

Various dependencies on the energy $E$ of the neutrino probes have been
assumed, in a phenomenological fashion, for the Lindblad decoherence
coefficient $\gamma = \gamma_{\rm Lnb} \left(\frac{E}{\mathrm{GeV}}\right)^n~,$
with
$n =0,2,-1$. The sensitivities in the work of \cite{lisi} from atmospheric
neutrinos (plus accelerator data~\cite{lisib}) at 90\% C.L. can be summarised
by the
following bounds on the parameter $\gamma_{\rm Lnb}$:
\begin{eqnarray}
\gamma_{\rm Lnb}& < & 0.4 \times 10^{-22} ~\mathrm{GeV}~, ~~ n = 0  \notag \\
\gamma_{\rm Lnb} & < & 0.9 \times 10^{-27} ~\mathrm{GeV}~, ~~ n = 2  \notag \\
\gamma_{\rm Lnb} & < & 0.7 \times 10^{-21} ~\mathrm{GeV}~, ~~ n = -1
\label{early}
\end{eqnarray}
Recently~\cite{lisi1}, updated values on these parameters, refered to 95\% C.L.,
have been provided by means of combining solar-neutrino and KamLand data
\begin{eqnarray}
\gamma_{\rm Lnb}& < & 0.67 \times 10^{-24} ~\mathrm{GeV}~, ~~ n = 0  \notag \\
\gamma_{\rm Lnb} & < & 0.47 \times 10^{-20} ~\mathrm{GeV}~, ~~ n = 2  \notag \\
\gamma_{\rm Lnb}& < & 0.78 \times 10^{-26} ~\mathrm{GeV}~, ~~ n = -1
\label{lisi1}
\end{eqnarray}
It should be remarked that all these bounds should be taken with a grain of
salt, since there is no guarantee that in a theory of quantum gravity $%
\gamma_{\rm Lnb}$ should be the same in all channels, or that the functional
dependence of the decoherence coefficients $\gamma$ on the probe's energy $E$
follows a simple power law. Complicated functional dependencies $\gamma (E)$
might be present.

We shall come back to these bounds in the discussion section of the
article,
when we compare the potential of upcoming neutrino data, with energies of
order of tens of GeV, from CNGS facility. We also
investigate the sensitivity of the experiments at J-PARC. The J-PARC
beam operates at rather lower energies comparing to CNGS, however the fact that the maximum
of oscillation in the spectrum will be measured by T2K experiment, allows to achieve a remarkable
sensitivity to those dapmping exponents with low power energy dependence, as
compared with~\cite{lisi,lisi1}. The CNGS is very sensitive to the $E^2$
dependent case despite the fact that the spectrum of atmospheric neutrinos used
in~\cite{lisi} spans a wide range of energy which extends to 100-1000~GeV. The
$E^2$ dependence, for
instance, could characterise Cauchy-Lorentz stochastic models of space time
foam.

In \cite{bmsw} a three generation Lindblad decoherence model of neutrinos
has been compared against all available at the time experimental data,
taking into account the recent results from KamLand experiment~\cite{kamland}
indicating spectral distortions.

\begin{figure}[tb]
\begin{centering}
\includegraphics[width=7.0cm]{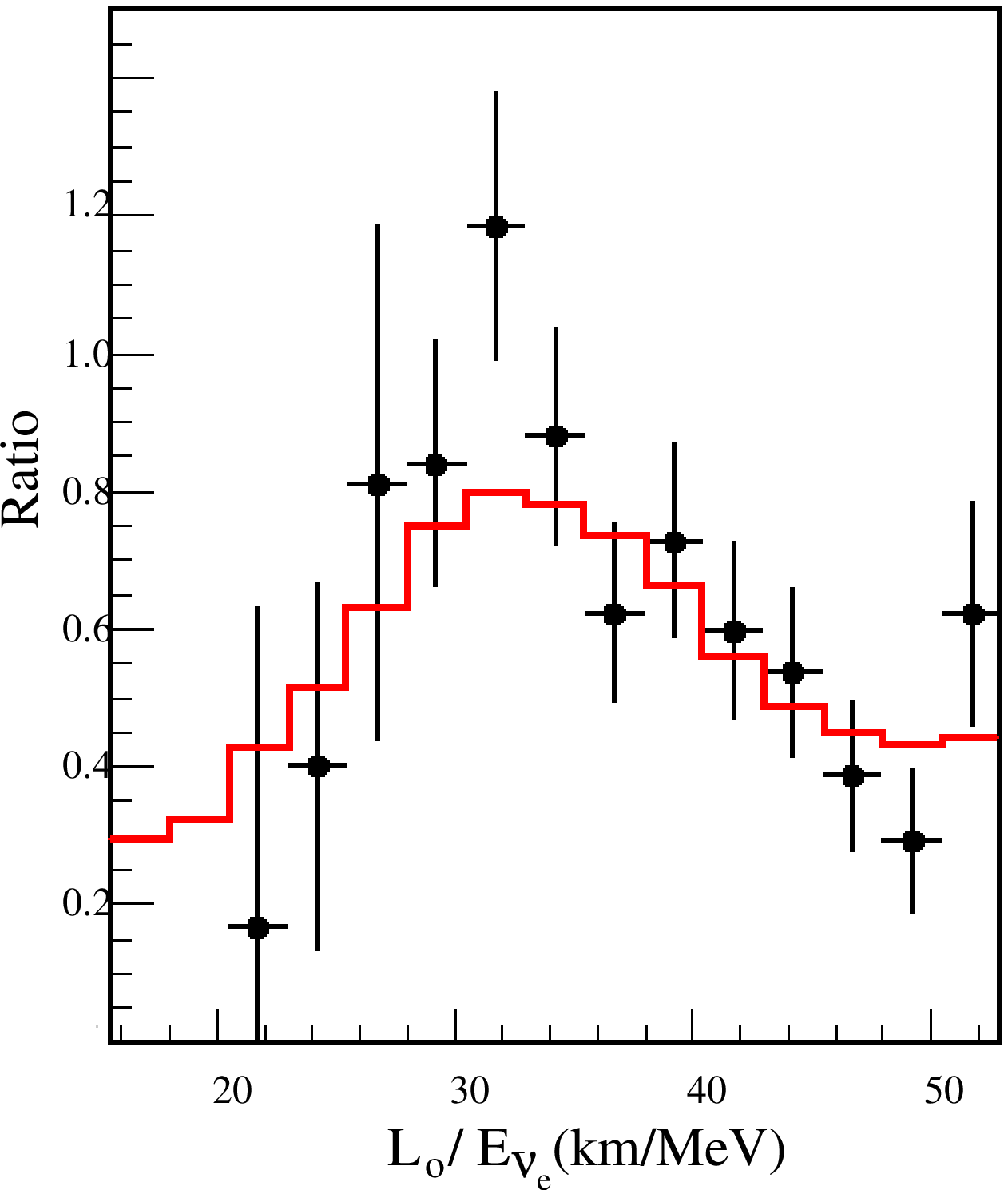}
\includegraphics[width=8.0cm]{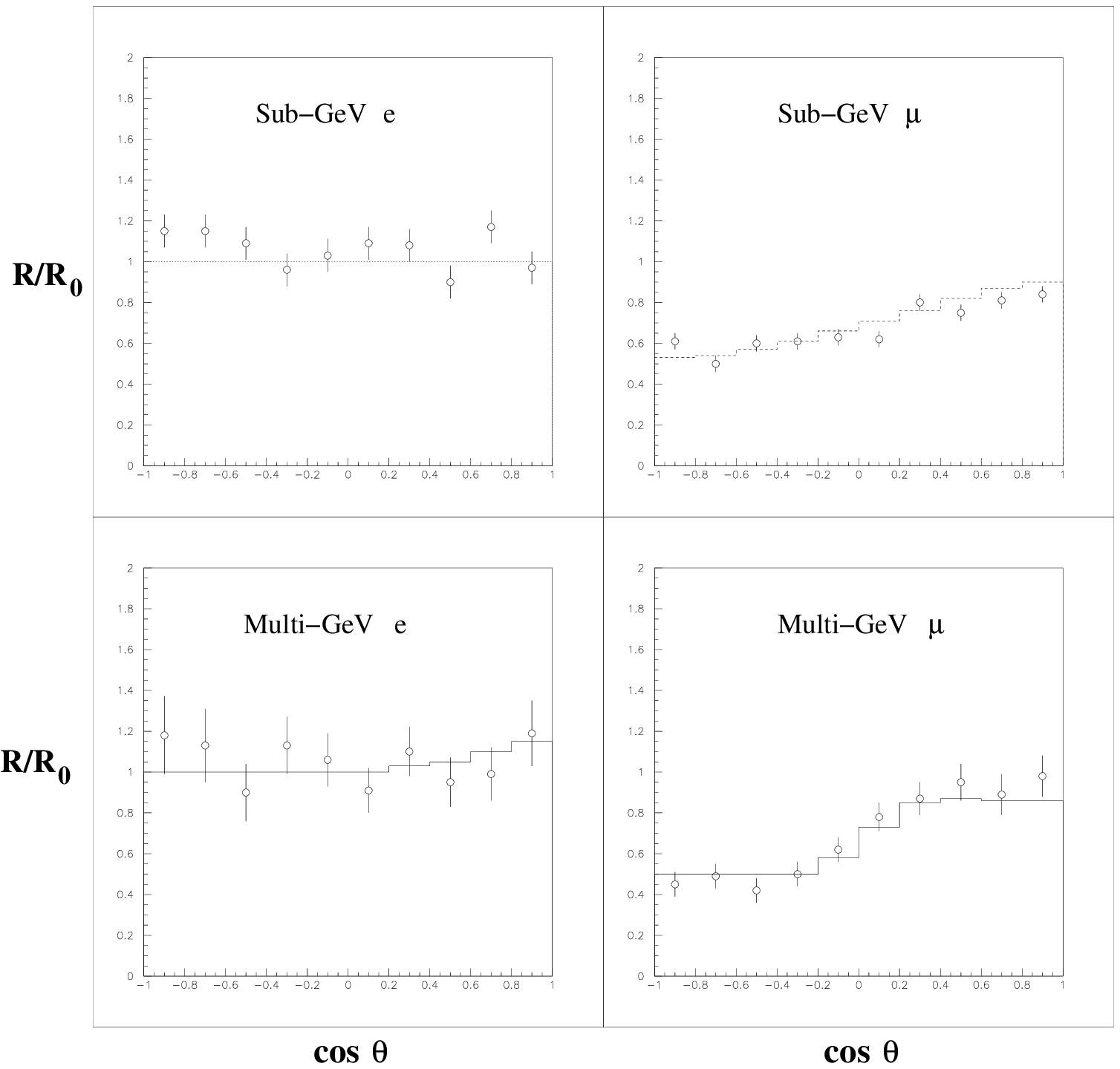}
\end{centering}
\caption{\protect\underline{Left}: Ratio of the observed $\overline{\protect%
\nu}_{e}$ spectrum to the expectation versus $L_{0}/E$ for our decoherence
model. The dots correspond to KamLand data. \protect\underline{Right}:
Decoherence fit. The dots correspond to SK data.}
\label{fig1}
\end{figure}

The results are summarized in Fig.~\ref{fig1}, which demonstrates the
agreement (left) of the model with the KamLand spectral distortion data~\cite%
{kamland}, and the best fit (right) for the Lindblad decoherence model used
in ref.~\cite{bmsw}.

The best fit has the feature that only some of the oscillation terms in the
three generation probability formula have non trivial damping factors, with
their exponents being \textit{independent} of the oscillation length,
specifically~\cite{bmsw}. If we denote those non trivial exponents as $%
\mathcal{D}\cdot L$, we obtain from the best fit of \cite{bmsw}:
\begin{align}
\mathcal{D}=- \frac{\;\;\; 1.3 \cdot10^{-2}\;\;\;} {L},  \label{special}
\end{align}
in units of 1/km with $L=t$ the oscillation length. The $1/L$-behaviour of $%
\mathcal{D}_{11} $, implies, as we mentioned, oscillation-length independent
Lindblad exponents.

In \cite{bmsw} an analysis of the two types of the theoretical models of
space-time foam, discussed in section 2, has been performed in the light of
the result of the fit (\ref{special}). The conclusion was that the model of
the stochastically fluctuating media (\ref{2genprob}) (extended
appropriately to three generations~\cite{bmsw}, so as to be used for
comparison with the real data) cannot provide the full explanation for the
fit, for the following reason: if the decoherent result of the fit (\ref%
{special}) was exclusively due to this model, then the pertinent decoherent
coefficient in that case, for, say, the KamLand experiment with an $L
\sim180 $~Km, would be $|\mathcal{D}| = \Omega^{2} G_{N}^{2} n_{0}^{2}
\sim2.84 \cdot10^{-21}~\mathrm{GeV}$ (note that the mixing angle part does
not affect the order of the exponent). Smaller values are found for longer $%
L $, such as in atmospheric neutrino experiments. The independence of the
relevant damping exponent from the oscillation length, then, as required by (%
\ref{special}) may be understood as follows in this context: In the spirit
of \cite{barenboim}, the quantity $G_{N} n_{0} = \xi\frac{\Delta m^{2}}{E}$,
where $\xi\ll1$ parametrises the contributions of the foam to the induced
neutrino mass differences, according to our discussion above. Hence, the
damping exponent becomes in this case $\xi^{2} \Omega^{2} (\Delta m^{2})^{2}
\cdot L /E^{2} $. Thus, for oscillation lengths $L$ we have $L^{-1}
\sim\Delta m^{2}/E$, and one is left with the following estimate for the
dimensionless quantity $\xi^{2} \Delta m^{2} \Omega^{2}/E \sim1.3
\cdot10^{-2}$. This implies that the quantity $\Omega ^{2}$ is proportional
to the probe energy $E$. In principle, this is not an unreasonable result,
and it is in the spirit of \cite{barenboim}, since back reaction effects
onto space time, which affect the stochastic fluctuations $\Omega^{2}$, are
expected to increase with the probe energy $E$. However, due to the
smallness of the quantity $\Delta m^{2}/E$, for energies of the order of
GeV, and $\Delta m^{2} \sim10^{-3}$ eV$^{2}$, we conclude (taking into
account that $\xi\ll1$) that $\Omega^{2}$ in this case would be
unrealistically large for a quantum-gravity effect in the model.

We remark at this point that, in such a model, we can in principle bound
independently the $\Omega$ and $n_{0}$ parameters by looking at the
modifications induced by the medium in the arguments of the oscillatory
functions of the probability (\ref{2genprob}), that is the period of
oscillation. Unfortunately this is too small to be detected in the above
example, for which $\Delta a_{e\mu} \ll\Delta_{12}$.

The second model (\ref{gravstoch}) of stochastic space time can also be
confronted with the data, since in that case (\ref{special}) would imply for
the pertinent damping exponent

\begin{align}
& \left( \frac{(m_{1}^{2}-m^{2}_{2})^{2}}{2k^{2}} (9\sigma_{1}+\sigma
_{2}+\sigma_{3}+\sigma_{4})+ \frac{2V\cos2\theta(m_{1}^{2}-m_{2}^{2})}{k}
(12\sigma_{1}+2\sigma_{2}-2\sigma_{3}) \right) t^{2}  \notag \\
& \sim1.3 \cdot10^{-2}~.
\end{align}
Ignoring subleading MSW effects $V$, for simplicity, and considering
oscillation lengths $t=L \sim\frac{2k}{(m_{1}^{2}-m^{2}_{2})}$, we then
observe that the independence of the length $L$ result of the experimental
fit, found above, may be interpreted, in this case, as bounding the
stochastic fluctuations of the metric to $9\sigma_{1}+\sigma_{2}+\sigma_{3}+%
\sigma_{4} \sim1.3. \cdot10^{-2}$. Again, this is too large to be a quantum
gravity effect, which means that the $L^{2}$ contributions to the damping
due to stochastic fluctuations of the metric, as in the model of \cite{ms}
above (\ref{gravstoch}), cannot be the exclusive explanation of the fit.

The analysis of \cite{bmsw} also demonstrated that, at least as far as an
order of magnitude of the effect is concerned, a reasonable explanation of
the order of the damping exponent (\ref{special}), is provided by
Gaussian-type energy fluctuations, due to ordinary physics effects, leading
to decoherence-like damping of oscillation probabilities of the form (\ref%
{uncert}). The order of these fluctuations, consistent with the independence
of the damping exponent on $L$ (irrespective of the power of $L$), is
\begin{align}
\frac{\Delta E}{E} \sim1.6 \cdot10^{-1}
\end{align}
if one assumes that this is the principal reason for the result of the fit.

However, not even this can be the end of the story, given that the result (%
\ref{special}) pertains only to \textit{some} of the oscillation terms and
not all of them, which would be the case expected for the ordinary physics
uncertainties (\ref{uncert}). The fact that the best fit model includes
terms which are not suppressed at all calls for a more radical explanation
of the fit result, and the issue is still wide open.

It is interesting, however, that the current neutrino data can already
impose stringent constraints on quantum gravity models, and exclude some of
them from being the exclusive source of decoherence, as we have discussed
above. Of course, this is not a definite conclusion because one cannot
exclude the possibility of other classes of theoretical models of quantum
gravity, which could escape these constraints. At present, however, we are
not aware of any such theory. We would like now to revisit the above
constraints in upcoming neutrino data from the experiments at the CNGS and
J-PARC facilities. This is discussed in the next sections.

\section{The combined fit to Quantum-Gravity Decoherence Signatures\label%
{sec:4}}

In the previous sections we have discussed several theoretical models of
quantum-gravity-induced decoherence independently, assuming each time only
one dominant type of decoherence: (i) Lindblad-type, through the
representation of the quantum gravity space-time foam as a stochastic
medium, (\ref{2genprob}), (ii) stochastically fluctuating space-time
backgrounds, (\ref{gravstoch}), and (iii) induced decoherence-like
evolution, as a result of uncertainties in the energy and/or oscillation
lengths of the neutrinos, (\ref{uncert}).

The various types of decoherence can be mainly distinguished by the form of
their exponential damping factor, as far as the power of the oscillation
length $L$ in the exponent is concerned, and the associated energy
dependence~\cite{dump}. Model independent data fits should combine, in
general, the various types of decoherence-deformed oscillations, given that
dominance of one or the other type may not be necessarily a feature of a
quantum-gravity model.

It is the purpose of this section, and one of the main objectives of this
work, to establish the limit of sensitivity of CNGS and J-PARC beams, in a
model-independent way, to a simple parametrization of the above effects,
combined in a single model for oscillations between flavours $a,b =1...n$ of
the form:
\begin{align}
& \langle P_{\alpha\beta} \rangle= \delta_{\alpha\beta} -  \notag \\
& 2 \sum_{a=1}^{n}\sum_{b=1, a<b}^{n}\mathrm{Re}\left( U_{\alpha a}^{*}
U_{\beta a}U_{\alpha b}U_{\beta b}^{*}\right) \left( 1 - \mathrm{cos}%
(2\ell\Delta m_{ab}^{2}) e^{-q_{1}L - q_{2}L^{2}}\right)  \notag \\
& -2 \sum_{a=1}^{n} \sum_{b=1, a<b}^{n} \mathrm{Im}\left( U_{\alpha a}^{*}
U_{\beta a}U_{\alpha b}U_{\beta b}^{*}\right) \mathrm{sin}(2\ell\Delta
m_{ab}^{2}) e^{-q_{1}L - q_{2}L^{2}}  \notag \\
& ~\mathrm{with} \qquad\ell\equiv\frac{L}{4E}  \label{combinefit}
\end{align}
where $L \simeq t$ (in units of $c=1$) is the oscillation length. In general
one may parametrize the damping exponents by polynomials in $L$~\cite{dump}
of any degree, but parametrisations of degrees higher than 2 are not
favoured by the class of quantum-gravity decoherence models considered in
the literature so far~\cite{mavromatos,ms}, and reviewed above.

From (\ref{2genprob}), (\ref{gravstoch}), we observe that (\ref{combinefit})
is \emph{oversimplified} in that it ignores possible modifications of the
oscillation period, which do exist in various microscopic models as a result
of the decoherence or stochastic-medium effects. A complete theoretical
treatment requires solving the evolution equations for the reduced density
matrix of neutrinos in a combined situation involving simultaneously
stochastic fluctuations of the background space-time metric and interactions
(of Lindblad type) with a stochastically-fluctuating quantum-space-time
medium. This will be left for future work. However, for our purposes in the
current article, we note that it is a reasonable assumption that such
modifications to the oscillation period are suppressed as compared with the
ordinary oscillation terms, and as such the dominant, model-independent,
terms appear to be only the exponents of the damping factors. Concerning the
latter, we also observe from (\ref{2genprob}) that, in general, there are
slight differences among the various exponents accompanying the oscillation
terms in stochastic-medium models, which however are all of the same order
of magnitude, and hence the error one makes in assuming the simplifying
two-parameter ($q_{1},q_{2}$) damping decoherence form (\ref{combinefit}) is
negligible.

For our phenomenological purposes in this work, therefore, the only
important point to notice is that the parameters $q_{i}, i=1,2$ may be in
general energy dependent, expressing back-reaction effects of the (neutrino)
matter onto the fluctuating space-time. Following earlier treatments and
theoretical quantum-gravity-decoherence models~\cite{lisi,mavromatos} we
shall consider the following three cases of generic energy dependence of the
decoherence coefficients $q_{i},i=1,2$:
\begin{equation}
q_{i},~i=1,2 \quad\propto E^{n}, \quad n=-1, 0, 2
\end{equation}
where the reader should have in mind that in each case the pertinent
decoherence coefficient has the appropriate units, as being a dimensionful
quantity.

For our studies we use two sets of the one and two parametric models
covering the main variety of phenomenologies for quantum gravity induced
decoherence phenomena described by the expression (\ref{combinefit}). The
first set of the models under consideration concerns the presence of linear
Lindblad-type mapping operator in the equation for the evolution of the
density matrix for the pure neutrino quantum states \cite%
{lisi,lisib,Benatti:2001fa,Benatti:2000ph,ohlsson,mavromatos}. The
oscillation probabilities corrected for the decoherence effects with
different energy dependence in the exponentials read

\begin{itemize}
\item no neutrino-energy dependence
\begin{equation}  \label{no_E}
P_{\nu_{\mu}\rightarrow\nu_{\tau}}=\frac{1}{2}\sin^{2}(2\theta_{23})\left[
1-\exp(-5\cdot10^{9}\gamma_{0}L) \cos\left( \frac{2.54\Delta m^{2}}{E}%
L\right) \right]
\end{equation}

\item inversely proportional to the neutrino energy (\emph{e.g.} the case of
Cauchy-Lorentz type of stochastic foam~\cite{alexandre}, (\ref{dampingCL}))
\begin{equation}  \label{inv_E}
P_{\nu_{\mu}\rightarrow\nu_{\tau}}=\frac{1}{2}\sin^{2}(2\theta_{23})\left[
1-\exp(\frac{-2.54 \gamma_{-1}^{2}L}{E}) \cos\left( \frac{2.54\Delta m^{2}}{E%
}L\right) \right]
\end{equation}

\item proportional to the neutrino energy squared
\begin{equation}  \label{sqr_E}
P_{\nu_{\mu}\rightarrow\nu_{\tau}}=\frac{1}{2}\sin^{2}(2\theta_{23})\left[
1-\exp(-5\cdot10^{27}\gamma_{2}E^{2}L) \cos\left( \frac{2.54\Delta m^{2}}{E}%
L\right) \right]
\end{equation}
\end{itemize}

where $\gamma _{0}$, $\gamma _{-1}^{2}$ and $\gamma _{2}$ are measured in
eV, eV$^{2}$ length and eV$^{-1}$ respectively the mass square difference $%
\Delta m^{2}$, is measured in eV$^{2}$, the energy $E$, is measured in GeV;
and the path, $L$, is measured in km.

The second set of the models concerns the gravitational MSW stochastic
effect~(\ref{2genprob}) with linear and quadratic~(\ref{gravstoch}) time
dependent fluctuations of space-time foam described by
\begin{equation}
P_{\nu _{\mu }\rightarrow \nu _{\tau }}=\frac{1}{2}-\exp (-\kappa _{1})\frac{%
\cos ^{2}(2\theta _{23})}{2}-\frac{1}{2}\exp (-\kappa _{2})\cos \left( \frac{%
2.54\Delta m^{2}}{E}L\right) \sin ^{2}(2\theta _{23}),  \label{msw}
\end{equation}%
where the exponential damping factors are chosen as

\begin{itemize}
\item no energy dependence, with linear
\begin{equation}  \label{msw1}
\kappa_{1}=5\cdot10^{9}\alpha^{2}L\sin^{2} (2\theta);\ \kappa
_{2}=5\cdot10^{9}\alpha^{2}L(1+0.25(\cos(4\theta)-1))
\end{equation}

quadratic
\begin{equation}  \label{msw2}
\kappa_{1}=2.5\cdot10^{19}\alpha_{1}^{2}L^{2}\sin^{2} (2\theta);\
\kappa_{2}=2.5\cdot10^{19}\alpha_{1}^{2}L^{2}(1+0.25(\cos (4\theta)-1))
\end{equation}

and combined time evolution
\begin{align}  \label{msw1comb}
\kappa_{1}=(5\cdot10^{9}\gamma_{1}^{2}L+2.5\cdot10^{19}\gamma_{2}^{2}L^{2})%
\sin^{2} (2\theta);  \notag \\
\kappa_{2}=(5\cdot10^{9}\gamma_{1}^{2}L+2.5\cdot10^{19}%
\gamma_{2}^{2}L^{2})(1+0.25(\cos(4\theta)-1))
\end{align}

\item proportional to the neutrino energy, with linear
\begin{equation}  \label{msw3}
\kappa_{1}=5\cdot10^{18}\beta^2 EL\sin^{2} (2\theta);\ \kappa
_{2}=5\cdot10^{18}\beta^2 EL(1+0.25(\cos(4\theta)-1))
\end{equation}

quadratic
\begin{equation}  \label{msw4}
\kappa_{1}=2.5\cdot10^{28}\beta_{2}^{2} EL^{2}\sin^{2} (2\theta);\
\kappa_{2}=2.5\cdot10^{28}\beta_{2}^{2} EL^{2}(1+0.25(\cos (4\theta)-1))
\end{equation}

and combined time evolution
\begin{align}  \label{msw2comb}
\kappa_{1}=(5\cdot10^{18}\gamma_{1}^{\prime2}EL+2.5\cdot
10^{28}\gamma_{2}^{\prime2}EL^{2})\sin^{2} (2\theta);  \notag \\
\kappa_{2}=(5\cdot10^{18}\gamma_{1}^{\prime2}EL+2.5\cdot10^{28}\gamma
_{2}^{\prime2}EL^{2})(1+0.25(\cos(4\theta)-1))
\end{align}

\item proportional to the neutrino energy squared, with linear time
evolution
\begin{equation}  \label{msw5}
\kappa_{1}=5\cdot10^{27}\beta_{1}^{2} E^{2}L\sin^{2} (2\theta);\
\kappa_{2}=5\cdot10^{27}\beta_{1}^{2} E^{2}L(1+0.25(\cos (4\theta)-1))
\end{equation}
The energy and the path length in (\ref{msw1})-(\ref{msw5}) are measured in
GeV and km respectively, while the parameters in damping exponentials are
given in eV in respective power (see Table~\ref{table_cngs}  for details).
\end{itemize}

\section{Sensitivity of CNGS and J-PARC beams to quantum-gravity decoherence
\label{sec:cngs}}

In this section we study the expected sensitivity of the CNGS and J-PARC
beams to the quantum gravitational decoherence phenomena described by (\ref%
{no_E})-(\ref{msw5}), considering them as subdominant contributions to the
atmospheric oscillations effects.

Both CNGS and J-PARC are conventional neutrino beams where neutrinos are
produced by the decay of secondary particles (pions and kaons) obtained from
the collision of the primary proton on a graphite target. For the CNGS beam,
the protons come from the CERN-SPS facility with a momentum of 400 GeV/c
whereas in the case of the J-PARC~\cite{jparc} the protons are produced in
Tokay (Japan) and have a momentum of 40 GeV/c. The expected number of
protons on target per year at the nominal intensity is $4.5\times10^{19}$
and $1 \times10^{21}$ respectively for the CNGS and J-PARC beam and the
envisaged run length is 5 years in both cases.

Both beams will be used for long baseline neutrino experiments which,
starting from a $\nu_{\mu}$ beam, will search for neutrino oscillations. The
OPERA experiment will measure neutrino events on the CNGS beam using a 2
kton detector which relies on the photographic emulsion technique, located
at a baseline of 732 km; the first neutrino events were observed in August
2006~\cite{first_opera}.

The T2K experiment will use the J-PARC beam measuring neutrino events with
the Super-Kamiokande~\cite{sk_rev} detector (a water cerenkov detector with
an active volume of 22.5 kton) at a baseline of 295 km.

Although CNGS beam designed in a way to be optimised for the $\nu _{\mu
}\rightarrow \nu _{\tau }$ oscillation searches through the detection of $%
\tau $ lepton production in a pure $\nu _{\mu }$ beam there is also a
possibility to measure $\nu _{\mu }$ spectrum by reconstructing $\mu $ from
the charged current (CC) events caused by $\nu _{\mu }$. Moreover, for this
experiment, we can take advantage of high mean value for the energy of $\nu
_{\mu }$s which makes the exponential damping factors more pronounced for
some cases described in the previous section.

The number of $\mu$ is given by the convolution of the $\nu_{\mu}$ flux $%
d\phi_{\nu_{\mu}}/dE$ with the $\nu_{\mu}$ CC cross section on lead $%
\sigma_{\nu_{\mu}}^{\mathrm{CC}}(E)$, weigted by the $\nu_{\mu}\rightarrow%
\nu_{\mu}$ surviving probability $P_{\nu_{\mu}\rightarrow\nu_{\mu}}$, times
the efficiency $\epsilon_{\mu\mu}$ of muon reconstruction of a given
detector:
\begin{equation}  \label{mu_count}
\frac{dN_{\mu\mu}}{dE}=A_{\mu\mu}\frac{d\phi_{\nu_{\mu}}}{dE}%
P_{\nu_{\mu}\rightarrow\nu_{\mu}}\sigma_{\nu_{\mu}}^{\mathrm{CC}}(E)
\epsilon_{\mu\mu}
\end{equation}
where $A_{\mu\mu}$ is a normalisation factor which takes into account the
target mass and the normalisation of the $\nu_{\mu}$ in physical units. In
our study we assumed an overall efficiency $\epsilon_{\mu\mu}$ of $93.5\%$
for the OPERA experiment and of $90\%$ for the T2K one as stated in the
experiment proposals.

To estimate quantitatively the sensitivity of CNGS on $%
P_{\nu _{\mu }\rightarrow \nu _{\tau }}$ described by (\ref{no_E})-(\ref%
{msw5}), we simulated the theoretical spectra of the reconstructed $\nu
_{\mu }$ events for various values of damping parameters.
Since there is no near detector at the neutrino source the overall
normalisation of the un-oscillated neutrino flux cannot be controlled with the
precision
better than 20\%, therefore such a normalisation has been taken into account in our $\chi ^{2}$ analysis
to estimate the expected limits of sensitivity on the damping
parameters:
\begin{equation}
\chi ^{2}=\sum_{i}[x_{i}-aP_{i}]^{2}/\sigma _{i}^{2} + (1-a)^2/\tilde\sigma^2,  \label{chi2}
\end{equation}
where, $x_{i}$ is the expected number of $\nu _{\mu }$ CC events contained
in the i-th energy bin considering standard three flavour oscillation, $P_{i}
$ is the number of events in the i-th bin theoretically expected when some
decoherence parameters are considered and $\sigma _{i}$ represents the error
on the number of events in the i-th bin. The parameter $a$ represents the normalisation factor and the additional contribution $(1-a)^2/\tilde\sigma^2$ 
is related to the systematic uncertainty of the overall neutrino flux at the source ($\tilde\sigma =0.2$). 
This systematic unceartanty~\footnote{For our
estimations, we take the most
conservative 20\%
uncertanty of the overal flux normalization.} plays an important role
in the correct estimation of the sensitivity of the experiment especially when the shape of 
$\nu_{\mu }$ CC  events spectrum is not changed by the decoherence effects (i.e. exponents independent
on energy).

For the atmospheric parameters we
used $\Delta m^{2}=2.5\cdot 10^{-3}\mathrm{eV^{2}}$ and $\theta
_{23}=45^{\circ }$~\cite{sk}. The 3~$\sigma $ sensitivity on the damping
parameters is found by applying a cut on the value of the $\chi ^{2}$ of 9
and 11.83 respectively for 1 d.o.f and 2 d.o.f.

As the CNGS beam is designed to observe $\nu _{\tau }$, neutrinos will have
a high energy with a mean value of about 17 GeV. This represents an
advantage since it makes the exponential damping factors more pronounced for
some cases described in the previous section. For the OPERA experiment the
systematic uncertainty in the muon detection efficiency is negligible
compared to the statistical uncertainties, therefore the error $\sigma _{i}$
used in our analysis (Eq.~\ref{chi2}) represents the statistical error only.

\begin{figure}[t]
\begin{center}
\includegraphics[width=10.0cm]{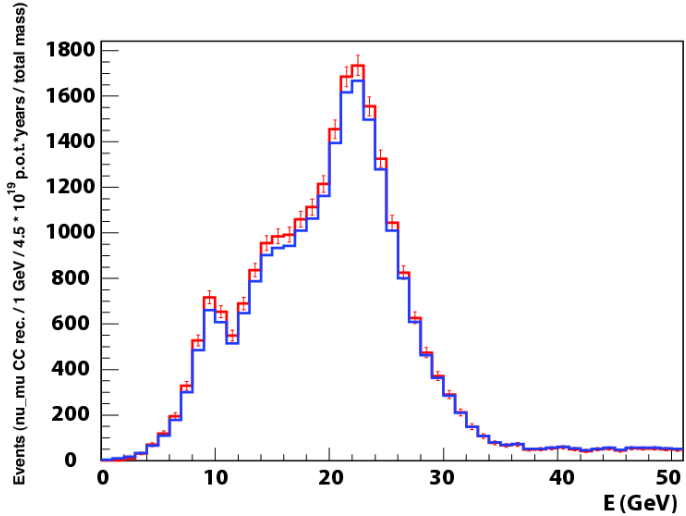}
\end{center}
\caption{The number of reconstructed $\protect\nu_{\protect\mu}$ CC events
in OPERA as a function of the neutrino energy with (blue line) and without
(red line with error bars) QG decoherence effect included in case of
inversely proportional dependence on neutrino energy. $3\sigma$ difference
between the expected and QG disturbed spectra is shown.}
\label{numu_spectrum}
\end{figure}

To generate the expected neutrino spectra of the CNGS beam measured by the
OPERA experiment we used a fast simulation algorithm described in~\cite%
{sim_nu} (see also Appendix~A for details). We present  in
Fig.\ref{numu_spectrum} a typical simulated
spectrum of the expected number of $\mu $ events including the effects of
decoherence (for the case of an inversely proportional dependence on
neutrino energy) as a subdominant suppression of the probability inferred
from the atmospheric neutrino experiment~\cite{sk}.

Our results for the sensitivity of CNGS to one parametric decoherence
damping exponentials in $P_{\nu _{\mu }\rightarrow \nu _{\tau }}$ are
summarised in second column of Table~\ref{table_cngs}. Also, for two parametric
fits~(\ref%
{msw1comb}) and~(\ref{msw2comb}), the 3~$\sigma $ CL sensitivity contours
are presented in Fig.~\ref{g1g2cngs} and Fig.~\ref{g1g2primcngs}.

\begin{figure}[t]
\begin{center}
\includegraphics[width=10.0cm]{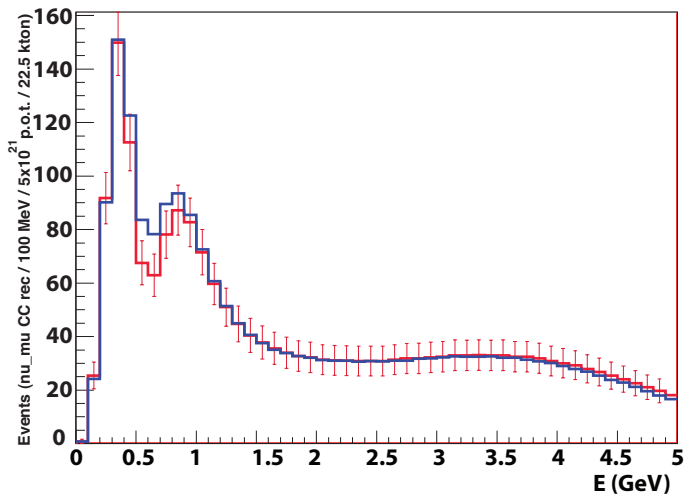}
\end{center}
\caption{The number of reconstructed $\protect\nu_{\protect\mu}$ CC events
in as a function of the neutrino energy with (blue line) and without (red
line with error bars) QG decoherence effect included in case of inversely
proportional dependence on neutrino energy. $3\sigma$ difference
between the expected and QG disturbed spectra is shown.}
\label{fig:T2K}
\end{figure}

\begin{figure}[t]
\begin{center}
\includegraphics[width=10.0cm]{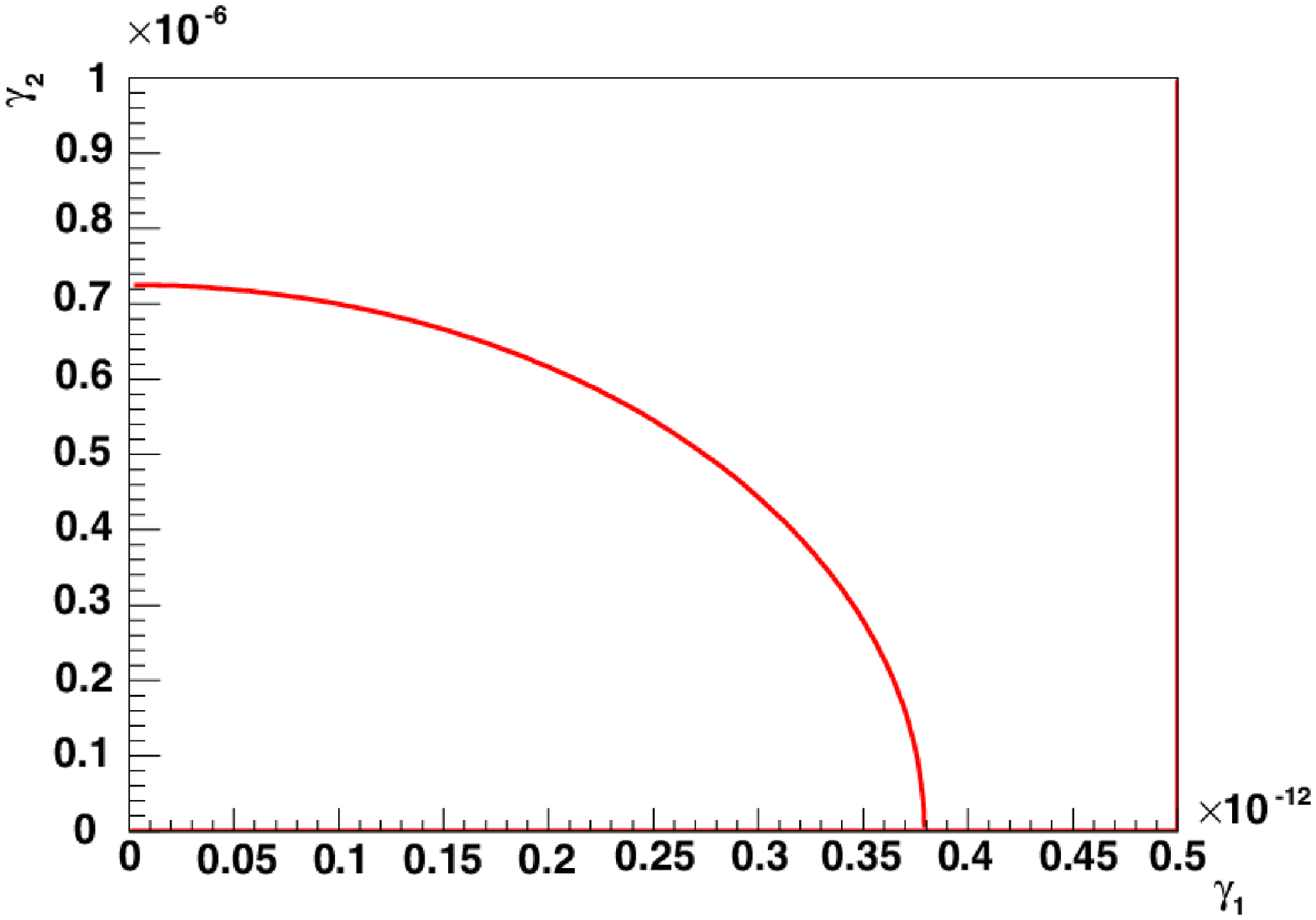}
\end{center}
\caption{The expected CNGS sensitivity contour at 3~$\protect\sigma$ CL, with
two
decoherence parameters contributing to the combined time evolution of
gravitational MSW (stochastic metric fluctuations) damping with no energy
dependence.}
\label{g1g2cngs}
\end{figure}

\begin{figure}[t]
\begin{center}
\includegraphics[width=10.0cm]{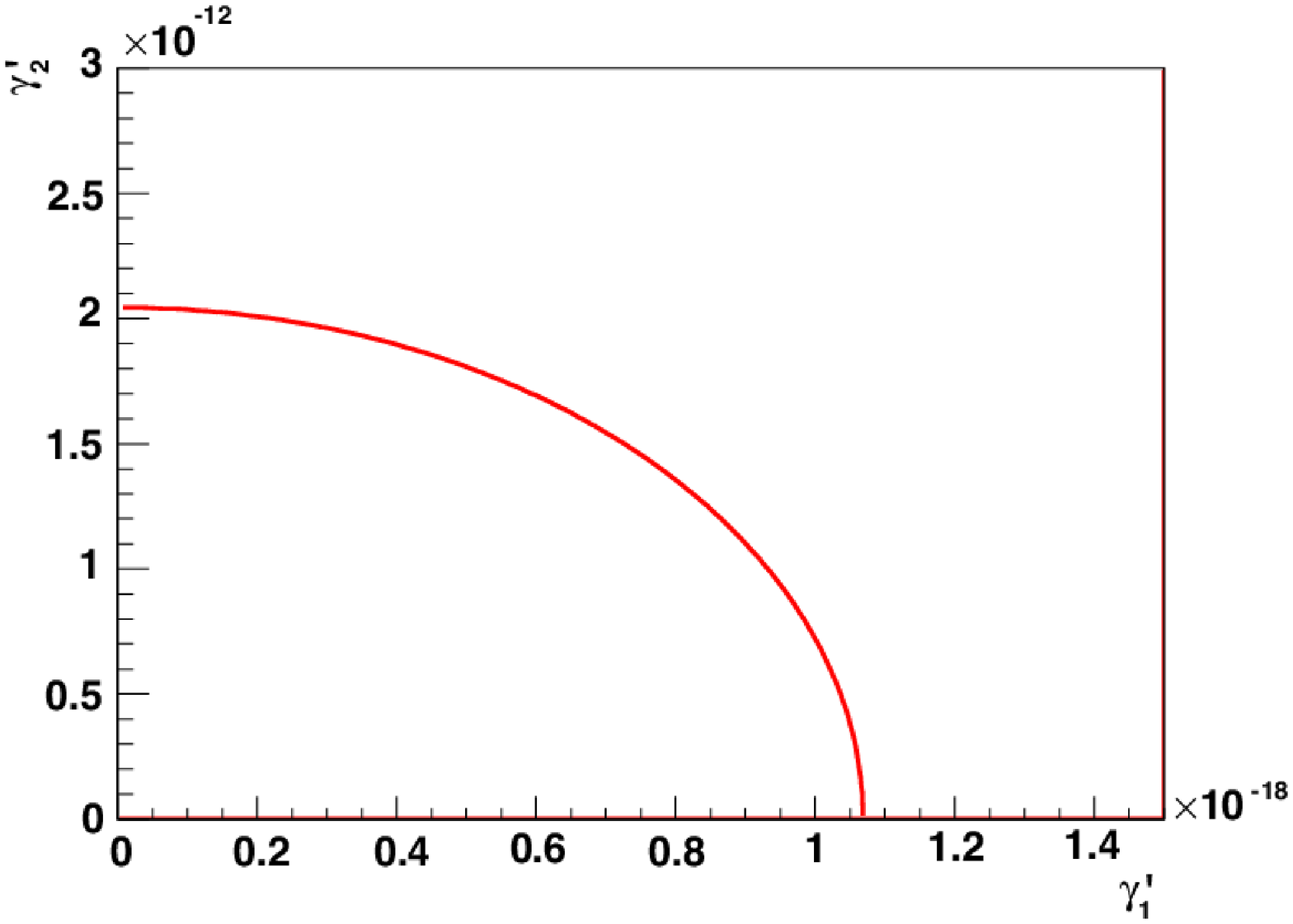}
\end{center}
\caption{The expected CNGS sensitivity contour at 3~$\protect\sigma$ CL, with
two
decoherence parameters contributing to the combined time evolution of
gravitational MSW (stochastic metric fluctuations) damping which are
proportional to the neutrino energy.}
\label{g1g2primcngs}
\end{figure}

Contrary to the OPERA experiment, the T2K experiment was designed to observe
$\nu _{e}$  and the mean energy is much lower: the maximum of oscillation at
the given baseline of 295 km corresponds to a neutrino energy of about 600
MeV and a narrow spectra at the selected energy will be obtained using the
so called off-axis technique~\cite{off-axis}. The spectrum
covers the region of the
first maximum of oscillation and this is a region where the QG effects could be
easily observed due to the small number of $\nu _{\mu }$ CC events expected
in case of no QG damping exponents, as it can be seen in Fig~\ref{fig:T2K}. 

The neutrino production at J-PARC beam is simulated
in GEANT environment, which takes into account the whole focusing system (horn
and reflectors), target and decay tunnel at J-PARC.
Protons are generated on target and through the decay of parent pions and kaons
the probability of neutrino at a selected location
is calculated and the spectra is obtained. We use the reconstructed neutrino
energy for single-Cherenkov-ring muon quasi-elastic (QE) and non-QE events. 
Of
course,
here the energy resolution plays an important role: for this reason we
introduced an energy smearing effect of 20\% in our analysis. This value 
takes into account the different energy resolution for the two kind of events and the
fact that QE events are the majority of the of muon neutrino events in the
detector.

\begin{figure}[t]
\begin{center}
\includegraphics[width=10.0cm]{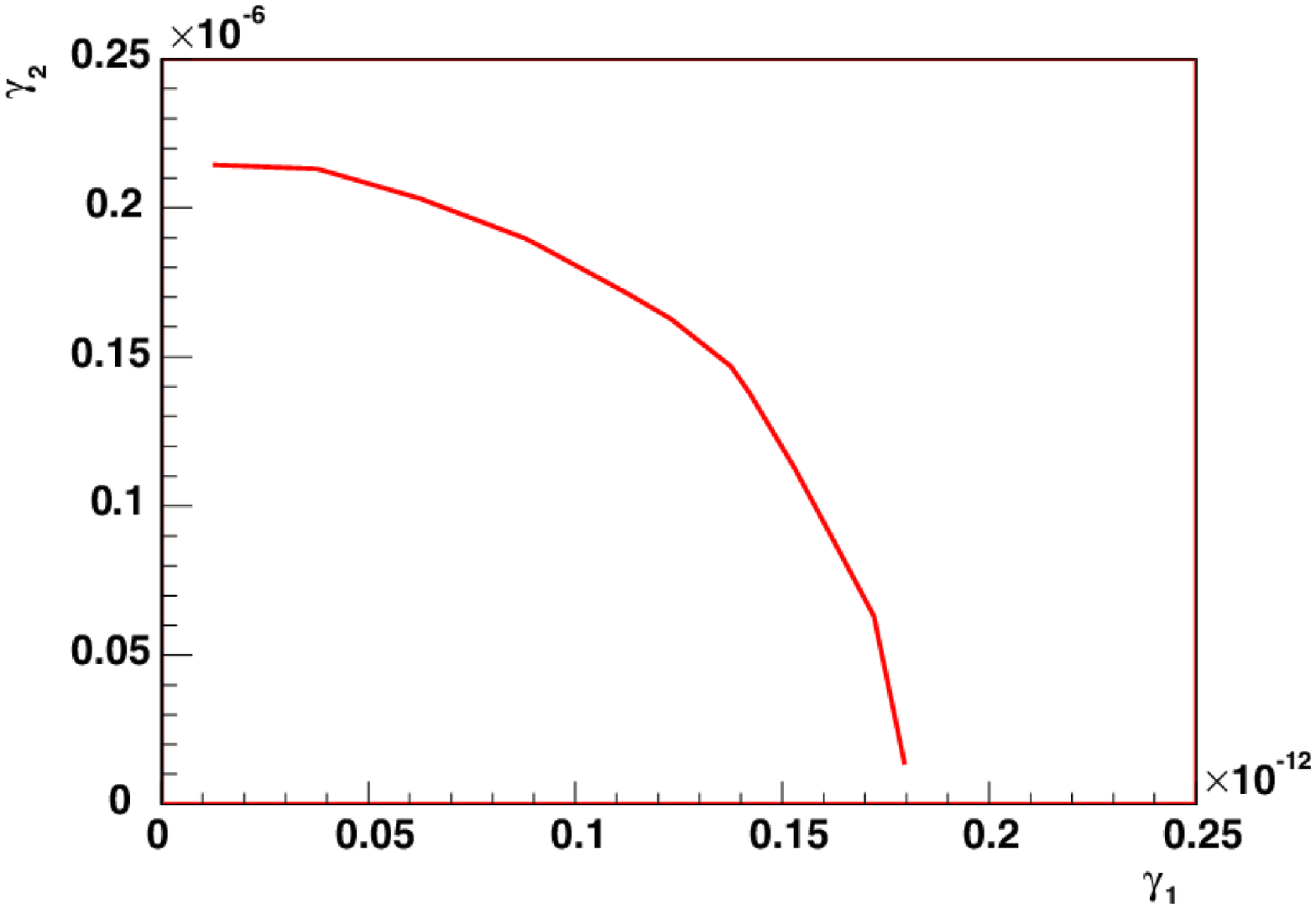}
\end{center}
\caption{The expected T2K sensitivity contour at 3~$\protect\sigma$ CL, with
two
decoherence parameters contributing to the combined time evolution of
gravitational MSW (stochastic metric fluctuations) damping with no energy
dependence.}
\label{g1g2t2k}
\end{figure}

\begin{figure}[t]
\begin{center}
\includegraphics[width=10.0cm]{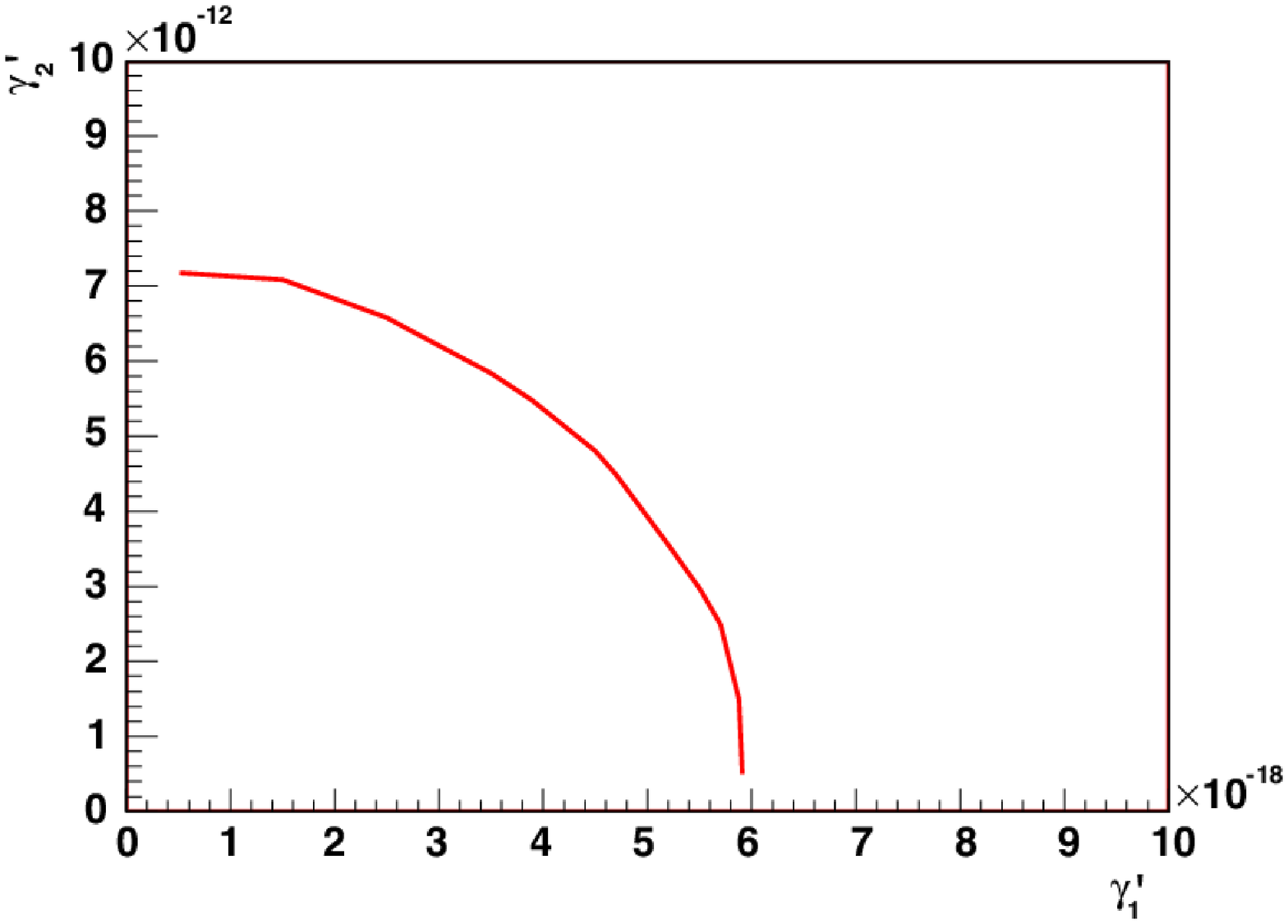}
\end{center}
\caption{The expected T2K sensitivity contour at 3~$\protect\sigma$ CL, with
two decoherence parameters contributing to the combined time evolution of
gravitational MSW (stochastic metric fluctuations) damping which are
proportional to the neutrino energy.}
\label{g1g2primt2k}
\end{figure}

Our results obtained using the same way of analysis quantified by~(\ref%
{mu_count}) and~(\ref{chi2}) for the sensitivity of T2K to one parametric
decoherence damping exponentials in $P_{\nu_{\mu}\rightarrow\nu_{\tau}}$ are
summarised in third column of Table~\ref{table_cngs}. Also, for two parametric
fits~(\ref%
{msw1comb}) and~(\ref{msw2comb}), the 3~$\sigma$ CL sensitivity contours are
presented in Fig.~\ref{g1g2t2k} and Fig.~\ref{g1g2primt2k}. Contrary to
CNGS, the J-PARC facility is equipped with a near detector which measures the
 un-oscillated muon spectrum with 5\% uncertanty in the absolute normalization
of the overal flux. However, to be conservative, we obtain our results
in 3rd and 4th column of Table~\ref{table_cngs} under the assumption
of 20\% uncertanty in the overal normalization of the spectrum. Since the main
effect is related to the maximal oscillation point in the spectrum, the overall
normalization is not as critical as in the case of CNGS fit.
{\
\begin{table}[b]
\begin{center}
{\scriptsize
\begin{tabular}{|c|c|c|c|}
\hline
Lindblad-type mapping operators &  CNGS & T2K & T2KK \\ \hline
\  &  &  &  \  \\
$\gamma_{0}$\ $[\mathrm{eV}]$\ ;\  ($[\mathrm{GeV}]$)& $2\times10^{-13}$\ ;\ 
 ($2\times10^{-22}$)& $2.4\times10^{-14}$\ ;\ ($2.4\times10^{-23})$
&
$1.7\times10^{-14}$\ ;\ ($1.7\times10^{-23}$)
\\
\  &  &  &\  \\
$\gamma_{-1}^{2}$\ $[\mathrm{eV^{2}}]$\ ;\  ($[\mathrm{GeV^{2}}]$)&
$9.7\times10^{-4}$\ ;\  ($9.7\times10^{-22}$)&
$3.1\times10^{-5}$\ ;\ ($3.1\times10^{-23}$) &
$6.5\times10^{-5}$\ ;\ ($6.5\times10^{-23}$) \\
\  &  &  &\  \\
$\gamma_{2}$\ $[\mathrm{eV^{-1}}]$\ ;\  ($[\mathrm{GeV^{-1}}]$)&
$4.3\times10^{-35}$\ ;\ ($4.3\times10^{-26}$) &
$1.7\times10^{-32}$\ ;\ ($1.7\times10^{-23}$) &
$3.5\times10^{-33}$\ ;\ ($3.5\times10^{-24}$) \\
\  &  &  &\  \\ \hline
Gravitational MSW (stochastic) effects &  CNGS & T2K & T2KK \\ \hline
\  &  &  &\  \\
$\alpha^{2}$ & $4.3\times10^{-13}~\mathrm{eV}$ & $4.6\times10^{-14}~\mathrm{eV}$
&
$3.5\times10^{-14}~\mathrm{eV}$ \\
\  &  &  &\  \\
$\alpha_{1}^{2}$ & $1.1\times10^{-25}~\mathrm{eV^{2}}$ &
$3.2\times10^{-26}~\mathrm{eV^{2}}$ & $6.7\times10^{-27}~\mathrm{eV^{2}}$\\
\  &  &  &\  \\
$\beta^{2}$ & $3.6\times10^{-24}$ & $5.6\times10^{-23}$ & $1.7\times10^{-23}$\\
\  &  &  &\  \\
$\beta_{2}^{2}$ & $9.8\times10^{-37}~\mathrm{eV}$ &
$4\times10^{-35}~\mathrm{eV}$ & $3.1\times10^{-36}~\mathrm{eV}$ \\
\  &  &  &\  \\
$\beta_{1}^{2}$ & $8.8\times10^{-35}~\mathrm{eV^{-1}}$ &
$3.5\times10^{-32}~\mathrm{eV^{-1}}$ &  $7.2\times10^{-33}~\mathrm{eV^{-1}}$\\
\  &  &  &\  \\ \hline
\end{tabular}
}
\par
{\scriptsize \vspace{0.3cm} }
\end{center}
\caption{Expected sensitivity limits at CNGS, T2K and T2KK to one parametric
neutrino decoherence
for Lindblad type and gravitational MSW (stochastic metric fluctuation)
like operators. }
\label{table_cngs}
\end{table}
}

The T2K experiment yields a better limit on the damping parameters only in
the case where the effect has no energy dependence or contains inversely
proportional to the neutrino energy exponent,
as expected given the low energy spectrum. In all the other cases, the
dependence on the baseline disfavours the short baseline of T2K with respect
to OPERA.

Another possibility to observe the effect on the T2K neutrino beam is to
select a longer baseline, namely to locate the detector at about 1000 km in
Korea. Studies of beam upgrades and a large liquid Argon detector of 100
kton in Korea were carried out~\cite{GlacierKorea} in the framework of CP
violation discovery. We considered this option, called T2KK, and studied the
possibility to constrain damping parameters in this case. The proposed
upgrade at 4 MW of the beam was taken into account which results into $%
7\times 10^{21}$ p.o.t. per year and a running time of 4 years was
envisaged. The efficiency $\epsilon _{\mu \mu }$ for the detector is assumed
to be $95\%$ and an energy smearing of $15\%$ is taken into account.

Our results for the sensitivity of T2KK to one parametric decoherence
damping exponentials in $P_{\nu _{\mu }\rightarrow \nu _{\tau }}$ are
summarised in fourth column of Table~\ref{table_cngs}. This configuration yields
better
results than the T2K experiment and results comparable to the OPERA
experiment.


\section{Discussion and Outlook \label{sec:discussion}}

It is instructive to compare the sensitivity limits presented in
Table~\ref{table_cngs} with those derived from the analysis
of
atmospheric neutrino data~\cite{lisi} obtained at Super-Kamiokande and K2K
experiments. One can transform the limits on the Lindblad type operators
presented in Table~\ref{table_cngs} to the notations~\form{early} of~\cite{lisi}
using the
following transformations: 
\begin{eqnarray}
\gamma_{\rm Lnb}& = & \gamma_0[{\rm GeV}], ~~ n = 0 \notag
\\
\gamma_{\rm Lnb} & = & \gamma_2[{\rm GeV^{-1}}]\times(\mathrm{GeV}^2) , ~~ n =2
\notag \\
\gamma_{\rm Lnb}& = & \gamma_{-1}^2[{\rm GeV^2}]/({\rm GeV}), ~~ n = -1
\label{lisi_transform},
\end{eqnarray}
so that the numbers of Table~\ref{table_cngs} in parentheses can be directly
compared with the bounds~\form{early} and~\form{lisi1}. 
 In particular, the bound obtained in~\cite{lisi} (see for
details~\form{early}) at 95\% C.L. on the
Lindblad type operators with no energy dependence is
close to the sensitivity estimated in our
analysis in case of T2K and T2KK simulations. Although, the CNGS estimation 
is about an order of magnitude weaker, one should stress that the current limit
is given at 99\% C.L. under the assumption of the most conservative
level of the uncertainty of the overall neutrino flux at the source. The bound
on the inverse energy dependence given in~\cite{lisi}~\form{early} is close
to the current
CNGS estimates. T2K and T2KK demonstrate an improvement. In spite of the fact that the
Super-Kamiokande data contains neutrino of energies up to $\sim$TeV, the
sensitivity one obtains at CNGS to the energy-squared dependent
decoherence is close, within an order of magnitude, to the bound~\form{early}
imposed by atmospheric neutrinos and surpasses T2K and T2KK sensitivity bounds
by $\approx 3$ and $\approx 2$ orders of magnitude respectively. The much less
uncertain systematics of CNGS compared to the atmospheric neutrino data will
make the expected bound more robust as soon as the upcoming data from OPERA will
be analysed. 
Moreover, our results are also
competitive with the sensitivity to the same Lindbland operators estimated
in~\cite{morgan} for ANTARES neutrino telescope, which is supposed to
operate at neutrino energies much higher than CNGS and J-PARC
experiments~\footnote{Although the sensitivity to the energy squired depending
damping exponent obtained in~\cite{morgan} is very close to our estimations
for CNGS, it is unclear why the authors of~\cite{morgan} claim a
remarkable improvement of this bound relative to the atmospheric
bound~\form{early}.}.

Assuming that the decoherence phenomena affect all particles in the same
way, which however is by no means certain, one might compare the results of
our analysis with bounds obtained using the neutral kaon system~\cite{cplear}%
. The comparison could be done for the constant (no-energy dependence)
Lindblad decoherence model. The main bound in~\cite{cplear} in such a case
reads $\gamma_{0}\le4.1\times10^{-12}$~eV, thus being about two orders of
magnitude weaker than the sensitivity forecasted in the present paper.

Finally, we compare the estimated sensitivity with the bounds obtained in~%
\cite{lisi1} using solar+KamLAND data. In principle, as in the case of the
neutral kaon system, a direct comparison is impossible, since the parameters
investigated here for the $\nu_{\mu}\rightarrow\nu_{\tau}$ channel need not
be the same for the $\nu_{e}\rightarrow\nu_{\mu}$ channel. However, again,
if these parameters are assumed to be roughly of equal size, then one can
see that the estimates of~\cite{lisi1} (\ref{lisi1}), which 
win essentially over the CNGS,  T2K and T2KK sensitivities only for the case of
inverse energy dependent decoherence, which strongly favours low neutrino
energies (\emph{%
e.g.} the case of Cauchy-Lorentz stochastic space-time foam models of \cite%
{alexandre} (\ref{dampingCL}), for which the current limit would bound, on
account of (\ref{inv_E}), the scale parameter $\xi $ of the distribution (%
\ref{cauchy}) to: $\xi < 5 \times 10^{-3}$ for neutrino-mass differences~%
\cite{lisi1} $|\mathrm{m}^2_{\mathrm{e}} - \mathrm{m}^2_\mu| = (7.92 \pm
0.71 )\times 10^{-5}$ eV$^{2}$). For the completeness, we mention that, our best
expected bound on the inverse-energy decoherence will imply, according
to~\form{lifetime} the bound on the $\nm$ life time
$\tau_{\nm}/m_{\nm}>3\times 10^{22}$~${\rm GeV}^{-2}$. 

The precise energy and length dependence of the damping factors is an
essential step in order to determine the microscopic origin of the induced
decoherence and disentangle genuine new physics effects from conventional
effects, which as we have seen in section \ref{sec:3}, may also contribute
to decoherence-like damping. Some genuine quantum-gravity effects, such as
MSW like effect induced by stochastic fluctuations of the space-time, are
expected to increase in general with the energy of the probe, as a result of
back reaction effected on space-time geometry, in contrast to
ordinary-matter-induced `fake' CPT violation and `decoherence-looking'
effects, which decrease with the energy of the probe~\cite{dump}. At
present, as one can see from the section \ref{sec:3}, the sensitivity of the
experiments is not sufficient to unambiguously determine the microscopic
origin of the decoherence effects, but according to our estimations of the
most plausible energy-length dependencies for the MSW like decoherence the
sensitivity of CNGS and T2K will improve the current limits by at least two
orders of magnitude and one would arrive at definite conclusions on this
important issue. Thus phenomenological analyses like ours are of value and
should be actively pursued when the data from OPERA and T2K will become
available. When the present paper was finished we got aware on a similar
analysis~\cite{t2k-t2kk} performed for J-PARC experiments  which agrees with
our results concerning T2K and T2KK.

In general, the characteristic energy dependencies of damping features are
very interesting to search for physics beyond the standard model. In some
cases, such damping signatures could be compensated by a shift of the
neutrino oscillation parameters, which means that given such a damping
effect, it is quite possible to obtain an erroneous determination of these
parameters. However, if the damping effects are strong enough, then an
establishment of effects beyond the standard neutrino oscillation scenario
will be possible. Once such a damping effect is established, it will be very
interesting to know from which non-standard mechanism it actually arises.
Given this  identification problem, we have found quite a low sensitivity
for the models with inverse energy dependence in the exponent, which means
that the damping effects of such kind are strongly correlated with the
standard neutrino oscillation parameters, i.e., it is difficult to
distinguish them from small adjustments in the oscillation parameters at
CNGS and T2K. However, damping signatures similar to energy dependence or
energy dependence squared can be very easily disentangled from the standard
oscillations, but it is difficult to distinguish them from each other.
Concerning different time dependencies in the exponents including the
combined signatures we have analysed, it can, in principle, be resolved if
there are two baselines, as applied for example in~\cite{t2k-t2kk}, and all the
other parameters are known. Also, for a
specific model, there may be relations among different $\gamma $'s in~(\ref%
{msw1comb}) and~(\ref{msw2comb}) like fits that actually imply much fewer
independent parameters.

\section{Acknowledgements}

The work of N.E.M. and S.S. is partially supported by the European Union
through the Marie Curie Research and Training Network \emph{UniverseNet}
MRTN-2006-035863.

\appendix

\section{Simmulation of neutrino beams}

A wide band neutrino beam is produced
from  the decay of mesons, mostly $\pi$'s and $K$'s. Mesons are
created by the  interaction of a proton beam into a needle shaped
target, they are  sign-selected and focused in the forward direction
by two large acceptance  magnetic coaxial lenses,  conventionally called
at CNGS (CERN) horn and reflector, and finally they are let to decay into an
evacuated tunnel pointing toward the detector position.

In case of positive charge selection, the beam content is mostly
$\nu_\mu$  from the decay of $\pi^+$ and $K^+$.  Small contaminations
of $\overline \nu_\mu$ (from the defocused $\pi^-$ and $K^-$) and $\nu_e$
(from three-body  decay of $K$'s and $\mu$'s) are present at the level
of few percent.

The neutrino fluxes for such a kind of beam are relatively easy to
predict~\cite{sim_nu}  once the secondary meson  spectra are known, because the
meson decay kinematics is well understood and  the geometry of the
decay tunnel is quite simple.

Uncertainty in the estimation of the neutrino fluxes could arise
because  secondary mesons are selected over a wide momentum range
and over a wide angular acceptance ($\simeq 20$ mrad).

Re-interactions of secondary  mesons in the target and downstream
material contribute to reduce the  neutrino fluxes and increase the
uncertainty in the calculations (mainly for the wrong sign and wrong
flavour contaminations).  They are generally minimized  using a target
made of a number thin rods of low Z material interleaved  with empty
spaces (to let the secondary mesons exit the target without
traversing too much material). In addition the amount of material
downstream  of the target (i.e. horn and reflector conductor
thickness) is kept  to the minimum.

The parameterization of the secondary meson production from protons
onto a thin target, proposed in~\cite{sim_nu}, is thus well suited to be
used in neutrino  beam simulations both because it extends its
prediction over a wide range of longitudinal and transverse
momenta and also because  the small fraction of tertiary production
from re-interactions in the target and downstream material can be
accounted for with the approximations described in~\cite{sim_nu}. 
A comparison of the neutrino flux prediction based on the
parameterization of~\cite{sim_nu} with some measured spectra is thus an
effective
estimator of the quality of the secondary mesons parameterization.
For this purpose, the parameterization~\cite{sim_nu} has been coupled with a
neutrino beam simulation program to be able to provide rapid and accurate
predictions of neutrino spectra at any distance (i.e. short and
long base line). The comparison has been performed both with already
published  data (CHARM II) and with predictions for the future CNGS
long-baseline neutrino beam generated with GEANT and/or
FLUKA based Monte Carlo programs.

The resulting code~\cite{code} is a stand-alone application developed on the
basis of parametrization~\cite{sim_nu} that allows to
vary and optimize all elements and the geometry (in 3-D)  of the beam
line providing  the results in terms of neutrino spectra and
distributions at large  distance with high statistics and in short
time.

The underlying idea is that in order to produce rapidly a neutrino
spectrum at large distance over a small solid angle at CNGS beam,
one has to force all the mesons  to decay emitting a neutrino, and
force all neutrinos to cross the  detector volume. A weight is then
assigned to each neutrino, proportional to the probability that
this process actually happened. In practice this method is implemented by
subdividing the simulation into  four subsequent steps, as described in detail
in~\cite{sim_nu}. These steps include: mesons production processes along
target, meson tracking in the neutrino beam-line, neutrino production
processes from mesons and neutrino production from muons. The weights
associated with every step is described in details in~\cite{sim_nu}.

\end{document}